\documentclass[letterpaper,11pt]{article}
\pdfoutput=1
\usepackage[usenames,dvipsnames]{xcolor}
\usepackage{graphicx}
\usepackage{epsfig,amsmath,amssymb,verbatim,mathrsfs}
\usepackage{bm}
\def\beq{\begin{equation}}
\def\eeq{\end{equation}}
\def\bea{\begin{eqnarray}}
\def\eea{\end{eqnarray}}
\newcommand{\prd}{Phys.Rev.D}

\usepackage[unicode=true,pdfusetitle,
 bookmarks=true,bookmarksnumbered=false,bookmarksopen=false,citecolor=Turquoise,linktoc=page,
 breaklinks=false,pdfborder={0 0 1},backref=false,colorlinks=true,pdfpagemode=FullScreen]
 {hyperref}

\newcommand{\gev}{\textrm{ GeV}}
\newcommand{\GeV}{\textrm{ GeV}}

\newcommand{\Deltamix}{ \Delta_{\textrm{mix}} }

\newcommand{\sr}{ s_{\textrm{r}} }
\newcommand{\vr}{ v_{\textrm{r}} }
\newcommand{\mur}{ \mu_{\textrm{r}} }

\newcommand{\sfone}{ s_{\textrm{u}_{1}} }

\newcommand{\vfzero}{ v_{\textrm{u}_{0}} }
\newcommand{\vfone}{ v_{\textrm{u}_{1}} }

\newcommand{\costwobetaeff}{ \cos2\beta_{\textrm{eff}}}

\newcommand{\sintwobetaeff}{ \sin2\beta_{\textrm{eff}}}
\newcommand{\sintwobetar}{ \sin2\beta_{\textrm{r}}}
\newcommand{\costwobetar}{ \cos2\beta_{\textrm{r}}}

\newcommand{\seff}{ s_{\textrm{eff}}}
\newcommand{\mueff}{ \mu_{\textrm{eff}}}
\newcommand{\veff}{ v_{\textrm{eff}}}
\newcommand{\mzeff}{ m_{Z,\textrm{eff}}}

\newcommand{\mzr}{ m_{Z,\textrm{r}} }

\begin{document}

\newcommand{\bit}{\begin{itemize}}
\newcommand{\eit}{\end{itemize}}

\baselineskip=17pt


\thispagestyle{empty}
\vspace{20pt}
\font\cmss=cmss10 \font\cmsss=cmss10 at 7pt

\begin{flushright}
\today \\
UMD-PP-012-021\\
\end{flushright}

\hfill

\begin{center}
\textsc{\LARGE Natural Islands for a 125 GeV Higgs in the scale-invariant NMSSM}
\end{center}

\vspace{15pt}

\begin{center}
{\large Kaustubh Agashe$\, ^{a}$, Yanou Cui$\, ^{a}$}, Roberto Franceschini$\, ^{a}$ \\
\vspace{15pt}
$^{a}$\textit{Maryland Center for Fundamental Physics,
     Department of Physics,
     University of Maryland,
     College Park, MD 20742, U.S.A.}
\end{center}

\vspace{5pt}

\begin{center}
\textbf{Abstract}
\end{center}
\vspace{5pt} {\small \noindent
}
We study whether a 125 GeV standard model-like Higgs boson can be
accommodated within the scale-invariant NMSSM in a way that is natural in all respects, i.e., not only is the
stop mass and hence its loop contribution to Higgs mass of natural size, but we
do not allow significant tuning of NMSSM parameters as well. We pursue as much as possible an analytic approach which gives clear insights on various ways to accommodate such a Higgs mass, while conducting complementary numerical analyses.
We consider both scenarios with singlet-like state being heavier and lighter than
SM-like Higgs.
With $A$-terms being small,
we find for the NMSSM to be
perturbative up to GUT scale,
it is not possible to get 125 GeV Higgs mass, which is true
even if we tune parameters of NMSSM.
If we allow some of the couplings to become non-perturbative
below the GUT scale, then the non-tuned option
implies that the singlet self-coupling, $\kappa$, is larger than the singlet-Higgs coupling, $\lambda$,
which itself is order 1. This leads to a Landau pole for these couplings close to
the weak scale, in particular below $\sim 10^{4}$ TeV.
In both the perturbative and non-perturbative NMSSM,
allowing large $A_{\lambda}, A_{\kappa}$  gives ``more room'' to accommodate a 125 GeV Higgs, but a tuning of these $A$-terms may be needed.
In our analysis we also conduct a careful study of the constraints on the parameter space from requiring global stability of the desired vacuum fitting a 125 GeV Higgs, which is complementary to existing literature. In particular, as the singlet-higgs coupling $\lambda$
increases, vacuum stability becomes more serious of an issue.

\vfill\eject
\noindent

\setcounter{tocdepth}{2}
\tableofcontents

\section{Introduction}

~~  Very recently, the ATLAS and CMS collaborations at the large hadron collider (LHC)
have discovered a resonance with a mass of around
125 GeV which is consistent with a standard model (SM) Higgs boson  \cite{lhc}.
Here, we assume
it {\em is} a SM-{\em like} Higgs.

In the SM, the quartic coupling for the Higgs and hence its mass is a
{\em free} parameter so that the above observation can be easily
``accommodated''. However, the SM is plagued by the Planck-weak hierarchy problem for which
weak-scale supersymmetry (SUSY) is perhaps the most studied solution.
In turn, in the minimal supersymmetric
SM (MSSM), the Higgs tree-level quartic coupling is given by the electroweak (EW) gauge coupling so that
there is actually a {\em prediction} for the Higgs mass.
The flip side is that the Higgs mass then has a
tree-level {\em upper} bound of $m_Z$, which is well below the above
observation
of 125 GeV.
It is well-known that the contribution of the superpartner of top quark (stop) at the loop-level can raise the
Higgs mass in the MSSM, but reaching 125 GeV this way requires stops to be as heavy as $\sim 5-10$
TeV or tuned, large $A_t$~\cite{Draper:2011aa,Hall:2011aa,Carena:2012zr,Wymant:2012vl}. The heaviness of
the stop results in reintroduction of fine-tuning of the EW scale.
This issue is embodied in the the relation
\begin{equation}
m_{Z}^{2} = -|\mu|^{2}-  \frac{m_{H_{u}}^{2}\tan^{2}\beta-m_{H_{d}}^{2}}{\tan^{2}\beta-1}\,, \label{mZmssm}
\end{equation}
that is one of the minimization constraints of the MSSM potential. Here the soft mass $m_{H_{u}}^{2}$ is largely determined
by the stop mass via RGE, resulting in $| m_{H_{u}}^{2} | \gtrsim 0.1\, m^{2}_{\tilde{t}}$ \cite{Giudice:2006sp}. Given the above
stop mass, such large value of $m_{H_{u}}^{2}$ needs to be canceled by other contributions at the right-hand side of eq.~(\ref{mZmssm}), in particular,
either against the
supersymmetric mass for Higgs doublets ($\mu$-term) or the other Higgs soft mass ($m_{H_{d}}^{2}$)
\cite{Dvali:1996ij,Csaki:2008dp,de-Simone:2011uq}.
At any rate the need for a cancellation of the large $m_{H_{u}}^{2}$ signals that a new ``little hierarchy'' needs to be explained in the MSSM.
This little hierarchy problem in MSSM can be quantified with a fine-tuning measure~\cite{Barbieri:1987fn}:

\begin{equation}
\Delta = \max_{\theta_{i}} \frac{d \log  m_{Z}^{2}} {d \log \theta_{i}} \, ,  \label{ewtuning}
\end{equation}
with $\Delta\lesssim 5$, i.e., less than $20\%$ tuning, conventionally taken as typical for a natural theory.

The MSSM also has another drawback, partly connected with the issue of having a cancellation in eq.~(\ref{mZmssm}) involving $\mu$. The issue is about why is the $\mu$-term close to the weak scale
as required phenomenologically: there is lower bound on it of $\sim 100$ GeV from chargino mass~\cite{The-DELPHI-Collaboration:-J.Abdallah:2003rr}, whereas
much larger values would require tuning in eq.~(\ref{mZmssm}).
An
attractive solution to this $\mu$-problem
is the scale-invariant NMSSM (for a review see \cite{Ellwanger:2009dp}). Here, an explicit $\mu$-term is forbidden in the superpotential, and instead is generated by the
vacuum expectation value (VEV) of an added SM gauge singlet, $S$, which is
coupled to Higgs doublets, $H_{u}$ and $H_{d}$, via the superpotential term
\begin{equation}
\lambda S H_{u}H_{d}\,.
\end{equation}
In turn, this VEV for the singlet is
driven by soft SUSY-breaking mass terms for singlet. As there are no explicit mass scales in the superpotential,
this model can be referred to as the scale-invariant NMSSM. This form of superpotential can be ensured by a $Z_3$-symmetry,
which can be extended to the soft SUSY breaking terms as well.
Thus it is often also referred to as the $Z_{3}$ invariant NMSSM~\footnote{The original $Z_3$ NMSSM typically suffers from the difficulty of simultaneously solving domain wall problem and tadpole problem which destabilizes electroweak scale~\cite{Abel:1995wk}. A simple way to resolve this issue is to assume a low-scale inflation such that the domain walls are significantly diluted~\cite{Ellwanger:2009dp}. Alternatively, a suitable R-symmetry can be imposed to constrain the form of explicit $Z_3$-breaking terms which may reconcile the tension of solving domain wall and tadpole problems~\cite{Abel:1996fk,Panagiotakopoulos:1998yw}.}.

As a {\em bonus} of the NMSSM, we get an extra tree-level quartic coupling
for the Higgs doublets from the same singlet-Higgs coupling $\lambda$ which solves the
$\mu$-problem. This effect in raising the Higgs mass relative to its MSSM prediction is well known and has been investigated at length in the literature (see for instance \cite{Ellwanger:2007kx,Barbieri:2006bg,Barbieri:2008yq}).
Here we re-visit this issue in the light of the discovery of a SM-like Higgs boson at 125 GeV.

Variations of the NMSSM
 may allow dimensionful terms in the superpotential.
In this work, we focus on the version of
the NMSSM with no
mass scales in superpotential so as to keep its other
 merit of solving the $\mu$-problem.
We assume that the stop contribution to Higgs mass
is small, as follows from natural $m_{\tilde{t}}, A_t$. This means that we aim at using tree-level
contributions to get close to 125 GeV, about 110 GeV or more, and
use the stop contribution to gain just the ``last'' 15 GeV or less in mass~\footnote{The estimate of 110 GeV lower limit will be detailed in the following and roughly corresponds to small $\tilde{t}_{L}-\tilde{t}_{R}$ mixing and the contribution of
a stop with mass of $500$ GeV which is a general upper bound from naturalness according to eq.~(\ref{ewtuning})  .
}.

\bigskip

Our goal is to see if it involves any tuning of parameters of the NMSSM to reach the required {\em tree}-level value for Higgs mass.
Such a suspicion is rather motivated.
For example, it is well known that  in a large region of the NMSSM parameter space, the lightest CP even state, $s_{1}$, is identified as the SM-like Higgs.  The mass of $s_1$ has an upper bound at tree-level
\begin{equation}
m_{s_{1}}^{2}\leq m_{Z}^{2} \cos^{2}2\beta + \lambda^{2} v^{2} \sin^{2}2\beta\,. \label{maxmh}
\end{equation}
If we prefer to preserve the conventional unification of the gauge couplings, which is a major success of MSSM, $\lambda$ is constrained to remain perturbative up to the GUT scale. This means that the value of $\lambda$ at the weak
scale which enters the above bound cannot be much larger than $0.6-0.7$  \cite{Masip:1998zr}. In turn,
even for such a maximal value of $\lambda$, the tree-level upper bound for the Higgs mass is very close to 110 GeV.
Given that Higgs mass of $110$ GeV or more at tree-level is our goal, it is clear that
we must actually {\em saturate} the upper bound in this case.
However, the  upper {\em bound} is not a generic value of the Higgs mass in the NMSSM, i.e.,
we expect that the model parameters must be arranged in a {\em specific} manner to fit the data in this case.
Here we shall investigate in detail what are the conditions that define the locus of maximal Higgs mass in this
region of the parameter space of the NMSSM.

A key quantity to identify the regions that lead to the largest Higgs boson mass (when it is the lightest CP-even state)
is the mixing of the doublet-like states with the singlet-like states. Indeed it has been known that the
extra quartic coupling of the NMSSM for the SM-like Higgs is not the ``end of the story"
since mixing with the singlet modifies the mass of the SM-like state \cite{Ellwanger:2007kx} and in general pushes it away from the value of the right-hand side of eq.~(\ref{maxmh}).
When the singlet-like scalar is heavier, as discussed so far,
its mixing with the doublet-like state results in a ``pull-down" of the mass of the latter.
The crucial point is that, even though the singlet-like scalar is heavier than the SM-like Higgs, it turns out that the
shift of the mass of the SM Higgs-like state
due to mixing with singlet state can still be significant, since this effect does not quite decouple even if singlet VEV is
large~\footnote{Although we might actually require
$\langle S\rangle\sim v_{\rm EW}$ in order to have a natural $\mu$-term, $\lambda \langle S\rangle$.
}.
As mentioned already, if the singlet-Higgs coupling $\lambda$ is to remain perturbative up to GUT scale, then
it turns out that the maximal tree-level SM-like Higgs mass~--~before negative mixing effect~--~can only be barely around or lower than our target of $110$ GeV. Thus one has  to work hard to
{\em minimize} the negative mixing effect
 on the SM-like Higgs mass in order to
obtain 125 GeV mass at loop-level.

There is actually a {\em different} possibility, namely, that the singlet-like scalar is actually lighter than the SM-like Higgs. In this case,
their mixing provides a ``push-up" effect, which raises the SM-like Higgs mass. Thus, the tree-level
mass, before the mixing is taken into account, can be {\em below} $110$ GeV, lessening the tension with perturbativity mentioned above for the other case.
However, we face a different potential worry: the state lighter than 125 GeV  necessarily has a component of
Higgs doublet so that it couples to the $Z$ boson and a bound from LEP2 applies~\cite{Schael:2006cr}. Thus, in this case, one has to work hard to make sure
that the singlet-like scalar is not too light or too strongly coupled to the $Z$ boson.

As we proceed with our above analysis, we distinguish two cases: the
``perturbative'' NMSSM that we mentioned above, which requires all the couplings do not hit a Landau pole till GUT scale,
and the non-perturbative NMSSM, where such singularity of the
%
%
couplings can appear below the GUT scale. The motivation is clear: the former
case can manifestly
maintain the successful gauge coupling unification of the
MSSM while in the latter case unification can be achieved only in specific UV completions~\cite{Harnik:2004fj, Birkedal:2004uq,Hardy:2012fk}.
Another useful ``axis'' of our exploration of the parameter space is the size of soft trilinear SUSY breaking terms
for Higgs doublets and singlet,  $A_{\kappa}$ and $A_{\lambda}$. We consider both cases where they are small and cases where they are large. This division clearly
is relevant
for some SUSY breaking mediation schemes that predict the $A$-terms to be small, for example, minimal gauge mediation.
Finally, as far as possible, we try to come up with analytical insights into these various issues.

The issue of the Higgs mass in the NMSSM has been studied recently, also in connection with the experimental finding of a 125 GeV SM-like Higgs boson. However our study is significantly different in a number of aspects with respect to the current literature. 
%
%
Some of them (see, for example, \cite{Hall:2011aa} and \cite{Ross:2012nr}) focus on more general singlet extensions of the MSSM, 
including 
the addition of explicit mass terms for the singlet and Higgs doublets.
Other studies such as \cite{Kang:2012qf,Arvanitaki:2011ck,Ellwanger:2011aa,Ellwanger:2012ad,Cao:2012hb,Cao:2012sf,Albornoz-Vasquez:2012yf} are mostly numerical and they lack a clear division into the above cases that we highlight.
%
%

Our results can be summarized as follows,
focusing on
%
%
the case with negligible $A$-terms.
For the perturbative case, we demonstrate that even if we allow tuning of parameters of NMSSM it
is {\em not} possible to reach 125 GeV mass:
for the sub-case here of pull-down, this limitation is due to a combination of the facts that the
singlet-Higgs coupling $\lambda$
is small to begin with and that the reduction in Higgs mass due to the mixing cannot be made small.
On the other hand,
for push-up, the failure to reach 125 GeV Higgs mass is due to the light singlet-like state typically being ruled out by LEP2.
In the
non-perturbative NMSSM, the singlet-Higgs coupling $\lambda$ and singlet self-coupling $\kappa$ can be larger.
Thus in this case the maximum tree level value itself can be well above 110 GeV.
Typically, one then {\em uses} the singlet-Higgs mixing
to get down from the above bound to 110 GeV at tree-level. In this way we completely avoid the tuning associated with requiring the pull-down mixing effect to be near vanishing
that plagued the perturbative case above.
Looking closer at the non-perturbative case it turns out that, if one wishes to avoid any kind of tuning of NMSSM parameters, then we
end up in a specific region, namely,
singlet-Higgs coupling, $\lambda$, is of order one and singlet self-coupling, $\kappa$, is larger.
%
%
%

One of the constraints in addition to $m_h \simeq 125 \GeV$ that we 
%
%
%
take into account, both analytically and numerically, 
is the
stability of the vacuum. Although this is a complicated issue to deal with, it is essential for the phenomenological viability of the model: the parameter choice that accommodates the right Higgs mass should correspond to a cosmologically stable vacuum. This issue, however, is either overlooked or only investigated in a partial way in most existing literature trying to explain $m_h\simeq \rm 125 \GeV$ within NMSSM. 
%

Here is the outline of the rest of the paper. We begin with reviewing the model and giving relevant formulae for our analysis. This  is followed by a general discussion of how to get to Higgs mass of 125 GeV, including relevant
issues such as perturbativity and vacuum stability. After this setting of the stage, we discuss in detail the viability
of the model in the
push-up scenario first and then the pull-down case, before concluding.

\section{The Model}
\label{sec:model}

The scale-invariant NMSSM is a very well-known extension of the MSSM which was originally
proposed to overcome the $\mu$ problem and can also serve to lift the tree-level Higgs mass. Nonetheless, in this section, for the sake of completeness and to fix our notation, we give some details
of the model relevant for our discussion. For a complete overview on the NMSSM we refer to~\cite{Ellwanger:2009dp}.

The superpotential of the model is
\begin{eqnarray}
W_{NMSSM}&=&\lambda \hat{S} \hat{H}_u \hat{H}_d+\frac{\kappa}{3} \hat{S}^3\,,
\end{eqnarray}
where $\hat{S}$ is a singlet chiral superfield and $\hat{H}_u = \left( \hat{H}_u^+, \hat{H}_u^0 \right)$ and
$\hat{H}_d = \left( \hat{H}_d^0, \hat{H}_d^-\right)$ are the MSSM Higgs doublets
(unhatted capital letters will indicate their complex scalar components).
The  soft-SUSY breaking Lagrangian is
\begin{eqnarray}
- {\cal L}_{ \rm soft } &=&m_{H_u}^2|H_u|^2+m_{H_d}^2|H_d|^2+m_S^2|S|^2+\lambda A_\lambda H_uH_dS+\frac{1}{3}\kappa A_\kappa S^3 \,.
\end{eqnarray}
Hence the complete scalar potential in this sector is:
\begin{eqnarray}
V_{ \rm scalar } & = & - {\cal L}_{ soft } + \sum_{ H_u, H_d , S } | F_i |^2 + \nonumber \\
 & & \frac{ g_1^2 + g_2^{ 2 } }{8} \left( | H^0_u |^2 +  | H^+_u |^2 -  | H^0_d |^2 -  | H^-_d |^2 \right)^2 +
\frac{ g_2^2 }{2} | H_u^+ H_d^{ 0 \; \ast } + H_u^0 H_d^{- \; \ast } |^2\, ,
\end{eqnarray}
where the last line is the MSSM $D$-term, $g_2 \approx 0.65$ and
$g_1 \approx 0.35$ denote the $SU(2)_L$ and $U(1)_Y$ gauge couplings,
and $F_i = \partial W / \partial \phi_i$ are the $F$-terms.

We expand the neutral scalar fields around the vacuum expectation values (VEVs) as follows (see, for example, \cite{Ellwanger:2009dp} ) :
\begin{eqnarray}
H_u^0 & = & v_u  + \frac{1}{ \sqrt{2} } \Big[ \left( h^0_v + i G^0 \right) \sin \beta +
\left( H^0_v + i A_v^0 \right) \cos \beta \Big] \,, \\
H_d^0 & = & v_d + \frac{1}{ \sqrt{2} } \Big[ \left( h_v^0 - i G^0 \right) \cos \beta - \left( H_v^0 - i A^0_v \right) \sin \beta
\Big]  \,, \\
S & = & s + \frac{1}{ \sqrt{2} } \left(  h_s^0+ i A_s^0 \right) \,,
\label{basis}
\end{eqnarray}
where $v = \sqrt{ v_u^2 + v_d^2 } \approx 174$ GeV, $\tan \beta = v_u / v_d$
and $G^0$ is the neutral  would-be Nambu-Goldstone boson.

The extremization conditions of the scalar potential are
\begin{eqnarray}
v_u \Big[ m_{ H_u }^2 + \mu^2 + \lambda^2 v_d^2 + \frac{ g_1^2 + g_2^{ 2 } }{4} \left( v_u^2 - v_d^2 \right) \Big] - v_d \mu B & = & 0 \,,\nonumber \\
v_d \Big[ m_{ H_d }^2 + \mu^2 + \lambda^2 v_u^2 + \frac{ g_1^2 + g_2^{2 } }{4} \left( v_d^2 - v_u^2
\right) \Big] - v_u \mu B & = & 0 \,, \nonumber \\
s \left( m_S^2 + \kappa A_{ \kappa } s + 2 \kappa^2 s^2
+ \lambda^2 v^2 - 2 \kappa \lambda v_u v_d \right) - \lambda v_u v_d A_{ \lambda } & = & 0 \, , \label{extremization}
\end{eqnarray}
where $\mu = \lambda s$ and $B = A_\lambda + \kappa s$.
We can use the above three equations in order to replace the parameters $m_{ H_u }^2$, $m_{ H_d }^2$ and $m_S^2$ by
the three VEVs.

The scalar Higgs states can be divided into a CP-even and a CP-odd sector. In fact, with the tree-level potential we consider there is neither explicit nor spontaneous CP violation in the Higgs sector~\cite{Romao:1986jy}.
The CP-even mass squared matrix in the above basis of $h^0_v$, $H^0_v$, $h_s^0$ is
\begin{equation}
{\cal M}^2 =
\left(
                    \begin{array}{ccc}
                      \lambda^2v^2\sin^22\beta+m_Z^2\cos^22\beta & 2rv^2\cot2\beta & 2\lambda^2sv-2v^2R \\
                      \cdot & -2v^2r+ 2\lambda \frac{\kappa s^2+A_\lambda s}{\sin2\beta} & -2Rv\cot2\beta \\
                      \cdot & \cdot & \kappa s(4\kappa s+A_\kappa)+\frac{v^2}{2s}A_\lambda\lambda\sin2\beta \\
                    \end{array}
                  \right)\,,
\label{CPeven}
\end{equation}
where $r\equiv\left(\frac{\lambda^2}{2}-\frac{m_Z^2}{2v^2}\right)\sin^22\beta$,
and $R=\frac{1}{v}\lambda(\kappa s+\frac{1}{2}A_\lambda)\sin2\beta$.

The basis used in eq.~(\ref{CPeven}) has the advantage that it shows clearly what state couples linearly to the SM
gauge bosons, such that it can be produced at LEP2 in association with a $Z$ boson.
In fact   $h^0_v$
is rotated in the same way as $v$, so that it is exactly the linear combination of
$H^0_u$
and $H^0_d$ which is responsible for the masses of the $W$ and $Z$ gauge bosons.
Thus
only $h^0_v$ has tree-level couplings to $WW/ZZ$, while $H^0_v$ and  $h_s^0$ do not.
This implies that only the component of $h^0_v$ in each CP-even mass eigenstate is relevant for the LEP2 limit.

Similarly,
to the extent that ATLAS and CMS
collaborations at the LHC have observed a SM-like Higgs at 125 GeV, in particular with
SM-like coupling to $WW/ZZ$, this state (denoted by $h$) should be predominantly
$h^0_v$, with small admixture of $H^0_v$, $h_s^0$.

The CP-odd scalar mass squared matrix in the above basis of $A_v^0$ and $A_s^0$ is given by
\begin{equation}
{\cal M}^2_A =
\left(
               \begin{array}{cc}
                 \frac{2\lambda s(A_\lambda+\kappa s)}{\sin2\beta} & \lambda v(A_\lambda-2\kappa s) \\
                 \cdot & \frac{\lambda v^2(A_\lambda+4\kappa s)\sin2\beta}{2s}-3\kappa A_\kappa s \\
               \end{array}
             \right) \,.
\label{CPodd}
\end{equation}
The Lagrangian is such  that we get a light CP-odd scalar
in the two approximate symmetry limits:
the $U(1)_R$ symmetric limit when the $A$-terms are small;
and the Peccei-Quinn symmetric limit when $\kappa$ and $A_{ \kappa }$ are small.
The light CP-odd scalars then correspond to
pseudo-Nambu-Goldstone bosons from {\em spontaneous} breaking of these symmetries.
If their mass is below about $10$ GeV,
there can be a constraint coming from $\Upsilon$-decay into these CP-odd scalars.
In general, the CP-odd states can also be produced in association with the CP-even scalars at LEP2.

Finally, there are physical charged Higgs bosons whose mass is given by
\begin{eqnarray}
m_{ H^{ \pm } }^2 & = & \frac{ 2 \mu B }{ \sin 2 \beta } + v^2 \left( \frac{ g_2^2 }{2} - \lambda^2 \right)\,.
\label{charged}
\end{eqnarray}

Although it is not directly relevant for our analysis,
we refer the reader to~\cite{Ellwanger:2009dp} for the mass matrix and the properties of the neutralinos and charginos.

Without loss of generality we take $\lambda>0$ while $\kappa$ and $\mu$ can have either signs. Meanwhile, in the examples shown in the following sections we take $k>0$ and $\mu>0$, as we do not expect the other choices of signs to result in any qualitative change of our findings.

\section{Getting SM-like Higgs of mass $125$ GeV}

As mentioned above, the 125 GeV SM-like Higgs discovered at the LHC is likely to be
predominantly $h^0_v$, with a small admixture of $H_v^0$, $h_s^0$.
Neglecting this mixing, its mass squared is given by ${\cal M}^2_{11 }$ in eq.~(\ref{CPeven}): the
$m_Z^2$ piece is the same as in the MSSM and the $\lambda^2$ piece results from the extra quartic of the NMSSM.
However, as mentioned in the introduction, the mixing of $h^0_v$ with $H^0_v$ and $h_s^0$ modifies the mass
from the above value, lowering (raising) if $h^0_v$ mixes with heavier (lighter) states. In fact the SM-like
Higgs mass can be written as:
\begin{eqnarray}
m_h^2 & = & m_Z^2 \cos^2 2\beta +  \delta m^2_{ h \; \rm loop}  + \nonumber \\
& & \lambda^2 v^2 \sin^2 2\beta + \delta m^2_{ h \; \rm mix} \,.
\label{schematic}
\end{eqnarray}

We further discuss below, in turn, each term in the above formula. The first line is the MSSM value, including the stop mixing effect for which we use the
approximate formula from eqs. (69-71) of \cite{Giudice:2006sp}:
\begin{eqnarray}
\delta m^2_{ h \; \rm loop } & = & \frac{ 3 \bar{m}^4_t }{ 2 \pi^2 v^2 }
\Big[  \ln \frac{ M_{ \tilde{t} }}{ m_t }+ \frac{ X_t }{4} + \frac{ \ln \frac{ M_{ \tilde{t} }}{ m_t } }{ 32 \pi^2 } \left( 3 \; m_t^2 / v^2 - 16 g_s^2 \right)
\left( X_t + 2 \ln \frac{ M_{ \tilde{t} }}{ m_t } \right) \Big]
\label{loop}
\end{eqnarray}
where the ``stop-mixing'' parameter $X_t$ is given by
\begin{eqnarray}
X_t & = & \frac{ 2 \left( A_t - \mu / \tan \beta \right)^2 }{ M_{ \tilde{t} }^2 } \Big[ 1 -
\frac{ \left( A_t - \mu / \tan \beta \right)^2 }{ 12 M_{ \tilde{t} }^2 } \Big],
\end{eqnarray}
$\bar{m}_t = m_t / \left(  1 + 4 \alpha_s / ( 3 \pi ) \right)$ is the $\overline{ \rm MS }$ top quark mass, $m_t$ being the pole mass,
$g_s$ is the QCD coupling
and $M_{ \tilde{t} }^2 \equiv m_{ \tilde{t}_1 } m_{ \tilde{t}_2 }$ is the geometric average of the
two stop mass eigenvalues.

We do not assume  ad-hoc large mixing in the stop sector which tends to increase the Higgs mass.
We also assume
500~GeV as a rough upper limit of stop mass $M_{ \tilde{t} }$ as it is favored by naturalness of weak scale:
 in particular, this corresponds to a fine-tuning of $\sim 20 \%$
if we assume the mediation scale of SUSY breaking of $\sim 10$ TeV, large $\tan \beta$ and small stop mixing 
(eq.~(6) of \cite{Papucci:2011wy})~\footnote{As emphasized in reference \cite{Hall:2011aa}, for $\lambda \sim 2$, even $1.5$ TeV stop mass can be natural. However,
for such large values of $\lambda$, the tree-level bound is already well above 125 GeV and so
a large loop contribution to Higgs mass from a heavy stop is not really needed anyway.}.
So we take typical values $\mu = 200$ GeV, $\tan \beta = 1.5$,
$M_{ \tilde{t} } \lesssim 500$ GeV and $A_t = 0$ in eq.~(\ref{loop}) and
find the stop loop contribution to $m_h$ can be up to $15$~GeV. In principle in the NMSSM there are further loop corrections involving the new couplings $\lambda$ and $\kappa$ \cite{Degrassi:2010lr}. For couplings around the bound imposed by perturbativity up to the GUT scale and stop masses saturating our naturalness upper bound  these corrections are in general subdominant to those stemming form the stop sector and we neglect them in the following. This may not be justified when one allows $\lambda$ and $\kappa$ to become non-perturbative. However in this case the Higgs mass can be larger at tree-level, hence loop corrections are in general less interesting to begin with. 
Given the 
loop corrections attainable from the stop sector, our goal here is to achieve a tree-level mass for the SM-like Higgs of $110$ GeV or more, up to $125$~GeV.
Of course if one would allow a heavier, hence less natural stop our analysis would be overly restrictive.
%

Moving onto the second line in eq.~(\ref{schematic}),
the first term here is the contribution characteristic of the NMSSM coupling $\lambda$. Notice that for a moderate or large $\lambda$, in the
NMSSM, $m_h$ is maximized at $\tan\beta\approx1$ in contrast to the case in the MSSM, where $m_h$ is maximized at $\tan\beta\rightarrow\infty$. Thus, we choose small $\tan \beta$  in our analysis.
As it will play an important role in what follows, we have introduced the notation  $\delta m^2_{ h \; \rm mix}$ in the second line
here to encode the  effect from the mixing of $h_v^0$ with $H_v^0$ and $h_s^0$.

For the discussion here, and for our analytic study in Sections \ref{pushupsection}  and \ref{pulldownsection}, we shall neglect the $h^0_v - H^0_v$ mixing and only deal with the $h^0_v - h_s^0$ mixing part of $\delta m^2_{ h \; \rm mix }$, reducing the 3-by-3 mixing problem into a simpler two states mixing problem. The justifications of this approximation are as follows.
Typically we have to choose parameters so that the $h^0_v - H^0_v$ and $h^0_s - H^0_v$ mixings are small effects.
The reason is that $H^0_v$ must be somewhat heavy, because $H^{0}_{v}$ is accompanied by a  charged Higgs of similar mass which is constrained by $b \rightarrow s \gamma$~\cite{The-BABAR-Collaboration:2012qf}.
As $H^{0}_{v}$ gets heavier its effect on $m_{h}$ decouples since the $h^0_v - H^0_v$ mass mixing term does not scale with $H^0_v$ mass, as shown in eq.~(\ref{CPeven}).
Furthermore in the NMSSM
small $\tan \beta$ is well motivated   and the
$h^0_v - H^0_v$ mixing actually vanishes at $\tan \beta = 1$.
In contrast, the effect of the $h^0_v - h_s^0$ mixing on SM-like Higgs mass
cannot be decoupled. The main reason  is that both the
$h^0_v - h_s^0$ mass term and $h_s^0$ mass scale with the VEV for $S$.

Similar to the $h^0_v - H^0_v$ mixing we neglect the $h_s^0 - H^0_v$ mixing since it affects the mass of the doublet like state only at higher order and it also vanishes at $\tan \beta = 1$.
We stress that we neglect $h^0_v - H^0_v$ and $h_s^0 - H^0_v$ mixing  only to give {\em analytical} arguments, of course, the full mixing is included in our numerical analysis.

As anticipated in the introduction, there are two possibilities for the mass of the SM-like Higgs boson as far as the effect of the mixing
effect in eq.~(\ref{schematic}) is concerned:
\begin{itemize}
\item The ``pull-down" case where  the singlet-like is heavier than SM-like Higgs,
such that the mixing reduces SM-like Higgs mass.\\
In this case, (i) ${\cal M}^2_{ 11}<{\cal M}^2_{ 33}$ and (ii) ${\cal M}^2_{ 11 } > ( 110 \;
\hbox{GeV} )^2$ such that after the pull-down mixing effect, we end up with a SM-like Higgs tree-level mass of
$110$
GeV.
Obviously, eq.~(\ref{CPeven})
then suggests that $\lambda \gtrsim 0.7$.
\item
The ``push-up" case where the singlet is lighter that the SM-like state,  so that the mixing increases
the SM-like Higgs mass.\\
By definition this implies that (i) ${\cal M}^2_{  11}  > {\cal M}^2_{33}$
and (ii) ${\cal M}^2_{  11} <  ( 110 \; \hbox{GeV} )^2$ since, after the effect of the mixing is included, we want to get to $110$ GeV at  tree-level.
Therefore, based on eq.~(\ref{CPeven}), we must have $\lambda \lesssim 0.7$.
\end{itemize}

\bigskip

Elaborating on the above discussion we are in position to outline the following argument.
If  the model has to be perturbative up to the GUT scale, then in the pull-down scenario, the region of the NMSSM parameter space that can give a SM-like Higgs mass as large as 125 GeV is not at all a generic one.
For the push-up case we shall indicate how we get a similar conclusion, though for a different reason.

\begin{figure}[t]
\begin{center}
\includegraphics[width=0.5\linewidth]{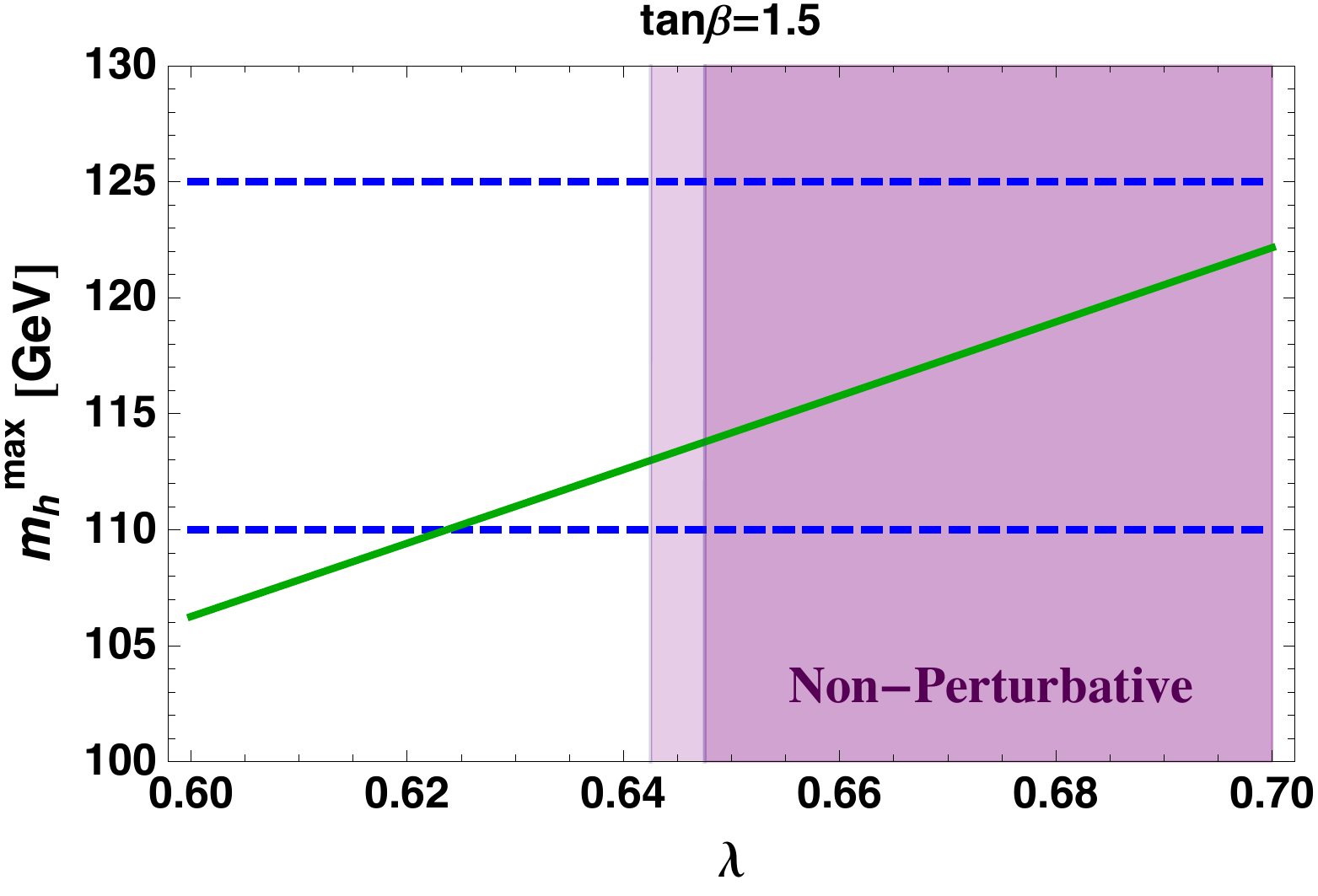}
\end{center}
\caption{The green solid line is the maximal mass of the SM-like Higgs boson for $\tan\beta=1.5$  from eq.~(\ref{schematic}) in the ``pull-down'' scenario. The two horizontal blue dashed lines are the reference values 125 GeV, that is about the observed mass of the SM-like boson, and 110 GeV, that is the minimal value of the tree-level mass to attain 125 GeV after the inclusion of radiative corrections from the stop, taking $M_{\tilde{t}}\lesssim500$~GeV. The light (dark) purple shaded regions mark the values of $\lambda$ that, according to the one-loop RGE, get larger than $\sqrt{2\pi}$, our definition for non-perturbative, below the GUT scale for $\kappa=0.1$ ($\kappa=0$) at the weak scale. To allow larger values of $\lambda$ to be compatible with perturbative unification we have
introduced extra matter in 4 copies of $5+\bar{5}$ of SU(5) at 5 TeV.  \label{mhmax} }
\end{figure}

If the coupling $\lambda$ must remain perturbative up to the GUT scale, it cannot exceed a generously estimated upper-bound at about 0.7~\footnote{In fact $\lambda$ is typically constrained to be below or around 0.6, as in general the need for a non-vanishing $\kappa$ reduces the maximal value of $\lambda$ compatible with perturbativity up to the GUT scale.}.
Even plugging in such large value of $\lambda$ in eq.~(\ref{schematic}) one finds that, in the pull-down scenario,
effect of the mixing must be minimal to keep the physical Higgs mass above 110~GeV at the tree-level.
This fact was
already mentioned in the introduction, but now we demonstrate it in Figure~\ref{mhmax} and it suggests that the parameters of the model must be arranged in a specific pattern such to have
\begin{equation}
\delta m_{ h \; \rm mix }^2 \simeq 0 \,.
\label{mhmix}
\end{equation}
This conspiracy of parameters seems rather unnatural, or can be taken as a suggestion for a peculiar relation that must come from a UV construction, and therefore we introduce a measure of ``tuning'' to quantify to what extent we are requiring cancellations among apparently unrelated parameters of the model.

In a fashion similar to what is done for the tuning of the EW scale (or $Z$ boson mass) as given in eq.~(\ref{ewtuning}), we introduce the tuning measure for the Higgs mass and expand the expression using eq.~(\ref{schematic}) and (\ref{CPeven}):
\begin{eqnarray}
\Deltamix &=& \max_{\theta_{i}} \frac{d \log \delta m^2_{ h \; \rm mix}}{d \log \theta_{i}}\nonumber  \\
&=&\max_{\theta_{i}} \frac{d \log \left(  m_{h}^{2} - {\cal M}^{2}_{11} \right)}{d \log \theta_{i}}\nonumber  \\
&=&\max_{\theta_{i}} \frac{d \log \left(  m_{h}^{2} - \lambda^{2}v^{2}\sin^{2}2\beta - m_{Z}^{2}\cos^{2}2\beta \right)}{d \log \theta_{i}}  \label{tuning}  \, ,
\end{eqnarray}
where $\theta_{i}=\lbrace \lambda,\kappa,A_{\lambda},A_{\kappa},\mu \rbrace$ are the various parameters of the NMSSM evaluated at the weak scale, where we compute the Higgs mass. This measure of tuning tries to capture how tightly related must  be the parameters of the NMSSM to attain 125 GeV for the mass of the SM-like Higgs state. A large value of $\Deltamix$ corresponds to a fine tuning among the values of two or more parameters.

In the following sections we will discuss both analytically and numerically what relations among the NMSSM parameters are implied by the recent LHC observation of a SM-like Higgs with
mass 125 GeV. Furthermore we will quantify using eq.~(\ref{tuning}) how natural are the identified regions of the parameter space. Altogether we will give the coordinates of the islands in the natural region of the NMSSM parameter space spotted by the LHC Higgs result.

For the push-up scenario the discussion follows a rather different path, but comes to a similar conclusion. The push-up scenario has a singlet-like state at the bottom of the spectrum. This state can be within the reach of LEP2 and to keep it unobservable in the dedicated searches of LEP2 experiments it must be significantly decoupled from the $Z$ boson, i.e.,
it must be mostly a pure singlet. In conflict with this is the fact that the lift of the Higgs mass, $\delta m^{2}_{h\; \textrm{mix}}$, goes to zero as the mixing angle vanishes.
Consequently in the push-up scenario the need for an almost pure singlet-like state at the bottom of the spectrum and the need to lift the Higgs mass are in tension. This picks out special regions of the parameter space where both the LHC and LEP2 constraints can be accommodated. In a spirit similar to that of the case of pull-down  we can quantify how ``special'' is the choice of the NMSSM parameters to attain such decoupled singlet in association with the needed lifting of the tree-level mass of the SM-like state. Since the issue is also about obtaining a small
$\delta m^2_{ h \; \rm mix}$, 
we can use the same measure of tuning introduced for the pull-down case as given in eq.~(\ref{tuning}).

\bigskip

In the following sections we will organize our analytical and numerical discussion on how to attain the mass of the SM-like Higgs at 125 GeV according to the push-up versus pull-down classification discussed above, paying particular attention to the size of $\Deltamix$.
Before getting into this detailed analysis,
we would like to comment on two other issues of the NMSSM, which become especially relevant for such a large mass of the SM-like Higgs. Namely, the stability of the desired vacuum
and the presence of Landau poles for the couplings, i.e., perturbativity of the model up to high scales.

\subsection{Perturbativity}

As we noted when commenting about  eq.~(\ref{schematic}),  $\tan \beta$ must be ``small" in order to make use of the extra quartic to increase the SM-like Higgs mass beyond its MSSM value. In fact for moderate or large $\lambda$ as used in this case, the tree-level upper bound on Higgs mass is itself maximized at $\tan\beta\approx1$.
However, such low $\tan \beta$
implies that the top Yukawa coupling is larger than $1$ already at the weak scale.  In addition to the danger that top Yukawa
hits Landau pole below GUT scale, such a choice, in turn, makes  $\lambda$ blow up faster~\footnote{On the other hand
negative stop 2-loop contribution is a bit smaller in size due to a cancellation between the now larger top Yukawa coupling and QCD coupling,
see eq.~(\ref{loop}).}. Therefore we shall choose $\tan \beta = 1.5$ as a ``representative" value for our analysis.

Also, in order to ameliorate the potential problem with Landau poles below GUT scales, we allow
the possibility for extra matter at low scales. Such matter should appear in complete representations of SU(5) so as to preserve
 gauge coupling unification, and tends to give larger MSSM gauge couplings in the UV. The gauge couplings, in
turn, prevent top Yukawa and singlet-Higgs couplings from blowing up too quickly.
%

\subsection{Constraints from unrealistic minima \label{extraminima}}
A set of important constraints on the values of the NMSSM parameters is that the ones fitting $m_h\simeq125 \GeV$ should correspond to a viable realistic vacuum. Basic checks include requiring it to be a minimum with all scalar masses squared being
positive. On top of this the desired vacuum there should be either a global minimum or a meta-stable minimum at cosmological time scale. Because of more parameters in NMSSM scalar sector, the check of vacuum stability is more sophisticated than in MSSM, in general with no neat analytic condition. Existing works  often overlook the issue of the global stability that, in particular, may become a
crucial one when $\lambda$ is large \cite{Kanehata:2011gf,Kobayashi:2012fj}.
In this work we take this issue seriously and conduct a thorough analysis on how vacuum stability may constrain the parameter space that fits $m_h\simeq 125 \GeV$. In this section we first give a {\em general} analytic formulation for such an analysis.
Then, we show a complete analytic study for a particular yet practically interesting limit
where $\mathcal{M}_{13}^{2}=0$, which, in turn, leads to $\delta m^2_{ h \; \rm mix } \sim 0$.
Recall that this is the condition which was identified as the key to attaining the desired Higgs mass while preserve perturbativity up to GUT scale: see the discussion after eq.~(\ref{schematic}).

Guided by this analytic understanding, in the numerical results contained in the following Sections \ref{pushupsection} and \ref{pulldownsection} we will impose
vacuum stability constraints on the parameter space that has an otherwise acceptable vacuum, i.e., which gives $m_h\simeq 125 $~GeV and, is compatible with the LEP bound when it is relevant. In the numerical studies the assessment of the global or local nature of the desired minimum is carried on by a global and unconstrained numerical search of possible extra minima of the potential. This must be contrasted with  the checks performed by standard tools \cite{nmssmtools} that evaluate analytical expressions for a (in general incomplete) set of the extremal points of the one-loop effective potential. Although limited by the precision of the numerical scan and by the fact that we use the tree-level potential, our method is in a sense more general as it can capture extra minima in any direction in field space. As we will see later, the requirement of absolute vacuum stability may rule out significant portions of the parameter space that are allowed by all other constraints. Our result gives a warning regarding the viability of some regions of parameters space and motivates a more detailed analysis at loop-level, that we leave for future work.

We start from stationary conditions of the NMSSM potential which determine the VEVs at extremum points, as given in eqs.~(\ref{extremization}). In general these 3 equations are coupled and the solution of each equation depends on the solution for the others. The equations are cubic in $s$, 
%
%
$v_{u}$ and $v_{d}$ therefore we count 
the following 
6 extremum solutions:
 \bit
 \item $(s=0, v=0)$,  $(s=0, v=\vfzero)$;
  \item $(s=\sr, v=0)$, $(s=\sr, v=\vr)$;
  \item $(s=\sfone, v=0)$, $(s=\sfone, v=\vfone)$,
  \eit
  where the ``r'' label denotes the desired quantities at the {\em realistic}  vacuum with $m_h\approx125\rm ~GeV$, $\vr^{2}  \equiv \mzr^2/g^2$, $g^2\equiv (g_1^2+g_2^2)/2$ and $\sr=\mur/\lambda$ corresponding to natural $\mur$.
 On the other hand, the ``u'' label denotes the nonzero quantities in
 extra {\it unrealistic}
  vacua. With $\lambda\neq0$ there is no runaway direction in NMSSM potential. Therefore in order to decide whether the desired vacuum is a global minimum, it is sufficient to check if $V_{\textrm{r}}$ is deeper than all other 5 extremum points. As already mentioned,
in order to keep our analytic study at a manageable level,
%
%
we
will work with
the {\em tree}-level potential. We are aware that, due to the importance of loop effects, this may not be accurate enough in some situations. In our numerical results presented later we shall consider the comparison of minima as ``unresolved'' when the value of the potentials at the two points  differ by less than 5\%~\footnote{This threshold of 5\% is our rough estimate of size of EW loop effects, which are expected to play a significant role in breaking
the tree-level degeneracy of the two vacua. Some numerical investigation of the impact of loop-corrections computed using NMSSMTools \cite{nmssmtools} confirm that our estimate is in the right ballpark. In fact the smallness of loop correction is expected once one chooses a natural stop mass.}.

 Applying stationary conditions on $H_u, H_d$ one obtains two practically interesting variables $\sintwobetaeff, \mzeff^2$ for each of the above solutions as functions of input parameters and the corresponding $\mueff$:
\bea
\sintwobetaeff&=&\frac{2(A_\lambda+\frac{\kappa}{\lambda} \mueff)\mueff}{m_{H_u}^2+m_{H_d}^2+2\mueff^2+\lambda^2\veff^2}\label{sin2beff} \,,\\
\mzeff^2&=&\frac{|m_{H_u}^2-m_{H_d}^2|}{\sqrt{1-(\sintwobetaeff)^2}}-m_{H_u}^2-m_{H_d}^2-2|\mueff|^2\label{mzeff}\,.
\eea
   Here in order to clarify things, we use the label ``eff'' as a general notation for output parameters from stationary conditions.
 Input parameters are those intrinsic to a model: $\lambda$, $\kappa$, $A_\lambda$, $A_\kappa$, $m_{H_u}^2$, $m_{H_d}^2$, $m_S^2$. Of course one set of $(\sintwobetaeff, \mzeff^2, \mueff)$ needs to be the values at the desired, realistic vacuum, which
as mentioned we will label with the label ``r''.

Fixing $\lambda,\kappa,A_{\lambda},A_{\kappa}$  we can  retrieve  the  mass squared model parameters from the quantities at realistic vacuum
\bea
  2m_{H_d}^2&=&\frac{2(A_\lambda+\frac{\kappa}{\lambda}\mur)\mur}{\sintwobetar}-2\mur^2-\lambda^2\vr^2\, \\\nonumber
  &+&
  \left(\frac{2(A_\lambda+\frac{\kappa}{\lambda}\mur)\mur}{\sintwobetar}-\mzr^2-\lambda^2\vr^2\right)\sqrt{1-(\sintwobetar)^2}\label{mhd2}\,,\\
  2m_{H_u}^2&=&\frac{2(A_\lambda+\frac{\kappa}{\lambda}\mur)\mur}{\sintwobetar}-2\mur^2-\lambda^2\vr^2\, \\\nonumber
  &-&\left(\frac{2(A_\lambda+\frac{\kappa}{\lambda}\mur)\mur}{\sintwobetar}-\mzr^2-\lambda^2\vr^2\right)\sqrt{1-(\sintwobetar)^2}\label{mhu2}\,,\\
  m_S^2&=&-\frac{1}{\sr}\left(\lambda^2\sr\vr^2+2\kappa^2\sr^3-\lambda\kappa \vr^2\sr\sintwobetar-\frac{1}{2}\lambda A_\lambda \vr^2\sintwobetar+\kappa A_\kappa \sr^2\label{ms2}\right) \,. \nonumber
\eea
Then one can (in principle) solve for other 5 extremum solutions $(s, v)$ and their corresponding $(\sintwobetaeff, \mzeff^2, \mueff)$. Finally we can plug all the relevant quantities into the potential  and compare the depths at different extrema. To achieve this goal we rewrite the potential as:
\beq
   V_{\rm min}=-\lambda^2\frac{\mzeff^4(\sintwobetaeff)^2}{16g^4}-\frac{\mzeff^4(\costwobetaeff)^2}{8g^2}+V^{S}_{\rm min}\label{potential} \,,
\eeq
where
\beq
   V^{S}_{\rm min}=\frac{\kappa^2}{\lambda^4}\mueff^4+\frac{2}{3}\frac{\kappa}{\lambda^3}A_\kappa\mueff^3+\frac{1}{\lambda^2}m_S^2\mueff^2 \,.
\eeq

\bigskip

Here a few comments are in order. Based on the first two terms in eq.~(\ref{potential}), among solutions with the
same   $\mueff$, the one with nonzero $\mzeff$ always has deeper potential. Therefore, among the 5 other extremum solutions we mentioned before, it is actually sufficient to just check if $(s=0, v=\vfzero)$ or $(s=\sfone, v=\vfone)$ has a deeper $V$ than desired vacuum. Therefore later in this section we focus on these two unrealistic minima.

The solution of $(s=0, v=\vfzero)$ is a special case, which is easily obtained from eqs.~(\ref{mzeff})~and~(\ref{sin2beff}):
      \bea
       \left.\sin2\beta\right|_{s=0}&=&0 \,,\\\nonumber
       \left. m_{Z} \right|_{s=0}&=&-2m_{H_u}^2 \,,\label{mu0sbmz}
      \eea
where we make the conventional ansatz $m_{H_u}^2<m_{H_d}^2$. This choice is motivated by radiative symmetry breaking, however choosing $m_{H_u}^2>m_{H_d}^2$ does not change the result.

On the other hand, the solution $(s=\sfone, v=\vfone)$ is hard to solve because this requires fully solving the 3 coupled cubic equations. However, there is a special limit where doublets can decouple from the stationary point equation of $S$, which greatly simplifies the process of obtaining the solution. To see this we recast the stationary point equation of $S$, the last one of eq.~(\ref{extremization}), in the form
  \begin{equation}
    2\kappa^2s^3+ A_\kappa s^2+m_S^2s+ \lambda v^2(\lambda s-\kappa s\sin2\beta-\frac{1}{2} A_\lambda\sin2\beta)=0\label{minisinglet}\,.
  \end{equation}
Apparently, doublet VEVs can be eliminated from eq.~(\ref{minisinglet}) when $v=0$ or the expression in the parenthesis vanishes.
Remarkably this requirement coincides with $\mathcal{M}_{13}=0$ which in turn implies the small singlet-doublet mixing eq.~(\ref{mhmix}). As we discussed before this is necessary to attain a 125~GeV mass for the SM-like Higgs while being compatible with perturbative unification.

In our following discussions we will first consider the competition between realistic vacuum and the generic $(s=0, v=\vfzero)$ solution given in eq.~(\ref{mu0sbmz}), in particular, the dependence on model parameters. Then we conduct a complete analysis for the simple special case mentioned above, i.e., when doublets decouple from stationary equation of $S$.
\subsubsection{$V_\textrm{r}$ vs. $V_\textrm{u}(s=0, v=\vfzero)$}
\label{s=0}

At the realistic vacuum, $\mueff=\mur$, $\mzeff=\mzr$, $\sintwobetaeff= \sintwobetar$. One can then plug the expression of $m_S$ in eq.~(\ref{ms2}) into eq.~(\ref{potential}), and simplify $V_\textrm{r}$ as:
\bea
V_{\textrm{r}}&=&-\lambda^2\frac{\mzr^4(\sintwobetar)^2}{16g^4}-\frac{\mzr^4(\costwobetar)^2}{8g^2}-\frac{\kappa^2}{\lambda^4}\mur^4-
\frac{\kappa}{3\lambda^3}A_\kappa\mur^3\label{vtrue}\\\nonumber
& &-\frac{1}{2}\mur \vr^2\left[2\mur-\sintwobetar(A_\lambda+\frac{2\kappa}{\lambda}\mur)\right] \,.
\eea
Notice that the second line in eq.~(\ref{vtrue}) vanishes in the limit when the doublets decouple from the stationary point equation
for $S$.

At the unrealistic extremum point  $(s=0, v=\vfzero)$, we clearly have $V^{S}_{\rm min}=0$ in eq.~(\ref{potential}). Thus,
the only relevant quantity here is $\mzeff$
which can be written in terms of quantities at the realistic minimum $\mur, \mzr, \sintwobetar$ using eqs.~(\ref{mhu2},\ref{mu0sbmz}). Finally, we get the following potential at this point:
\bea
V_{\textrm{u}}(s=0, v=\vfzero)\label{vmu0} &=& -\frac{\mzeff^4}{8g^2}\\\nonumber
&=&-\frac{1}{8g^2}\left[-\frac{2(A_\lambda+\frac{\kappa}{\lambda}\mur)\mur}{\sintwobetar}+2\mur^2+\lambda^2\vr^2  \right. \\ \nonumber
 &~& \left. +\left(\frac{2(A_\lambda+\frac{\kappa}{\lambda}\mur)\mur}{\sintwobetar}-\mzr^2-\lambda^2\vr^2\right)
  \sqrt{1-(\sintwobetar)^2}\right]^2\,.
\eea

So, the question of whether the realistic minimum is the global one or not reduces to a
comparison of the value of the potential in eq.~(\ref{vtrue}) and eq.~(\ref{vmu0}). We first make some general qualitative comments that  will be later confirmed by our quantitative studies.
With all the other parameters fixed, at a value of $\mur$ much larger than $m_Z$,
the unrealistic extremum tends to have deeper $V$  than $V_\textrm{r}$.
A general ``hint'' of this feature, that may also apply to the other unrealistic extrema,
comes from the fact that the $\mzeff^4$ term(s) in eq.~(\ref{potential}) give negative contribution to $V$.
At the realistic vacuum $\mzeff$ has the value 91~GeV, achieving which, according to eq.~(\ref{mzeff}), requires tuning when $\mur$ is large.
On the other hand, at the other extrema with non-zero $\mzeff$, the  size of $\mzeff$ is typically  $m_{Z,\textrm{u}}\sim m_{H_u}\sim\mu_\textrm{r}$.
Therefore $V_\textrm{u}$ gets a stronger negative push from larger $|-\mzeff^4|$ terms.
To be more specific to this extremum $(s=0,v=\vfzero)$
  we can see from eq.~(\ref{vtrue}) and eq.~(\ref{vmu0}) that, at large $\mur$
%
%
it is the
dominant negative contribution of $O(\mur^4)$ in the potential which tends to make $V_\textrm{u}$ deeper.
In fact  $V_\textrm{u}\supset-\frac{1}{2g^2}\left[1-\frac{\kappa}{\lambda}(1-\cot2\beta_\textrm{r})\right]^2\mur^4$ to be confronted with the contribution in the potential at the realistic minimum $V_{\textrm{r}}\supset-\frac{\kappa^2}{\lambda^4}\mur^4$.
With $\tan\beta_\textrm{r}\approx1.5, \lambda\approx0.7$ as a relevant example, we find the coefficient of $\mu_\textrm{r}^4$ term in $V_\textrm{u}$ is larger in magnitude (but negative) than in $V_\textrm{r}$, except for a small intermediate region of $0.5\lesssim\kappa\lesssim1$.

Furthermore one can study the dependence on $\lambda$ with all other parameters fixed. At larger $\lambda$, the
unrealistic extremum tends to have deeper $V$ than $V_\textrm{r}$. This can be seen by comparing the leading $\lambda$ terms in the two potentials. These are  $-\frac{\mzr^4(\sintwobetar)^2}{16g^4}\lambda^2\subset V_\textrm{r}$ and $-(\frac{1}{8g^2}(1-\sqrt{1-(\sintwobetar)^2})^2\vr^4\lambda^4+c\mur^2\vr^2\lambda^2)\subset V_\textrm{u}$  where the $c-$term has an $O(1)$ coefficient depending on $A_\lambda, \kappa$. 
Clearly,
the leading $O(\lambda^4)$ term gives a significant negative contribution to $V_\textrm{u}$ which is absent in $V_\textrm{r}$.

We can also study the dependence on $\kappa$ with all other parameters fixed. At larger $\kappa$, the unrealistic extremum tends to be shallower than $V_\textrm{r}$. This can be seen by comparing coefficients of the leading $\kappa$ dependent term of $O(\kappa)$ at large $\kappa$. These are $-\frac{\kappa^2}{\lambda^4}\mur^4\subset V_\textrm{r}$ and $-\frac{(1-\cot2\beta_\textrm{r})^2}{2g^2\lambda^2}{\kappa}^2\mur^4\subset V_\textrm{u}$.  With $\tan\beta_\textrm{r}\approx1.5, \lambda\approx0.7$ as an example, $V_\textrm{r}$ gets a larger negative contribution at larger $\kappa$.

\subsubsection{A complete vacuum analysis at doublet-singlet decoupling limit}
\label{complete}

 As mentioned earlier, it is much easier to solve for all extrema in the NMSSM potential in the  limit
 of decoupling doublet from singlet.
 Strikingly, this is also the interesting limit for obtaining $m_h\simeq 125$~GeV while being compatible
 with perturbative unification and LEP limits.
Here we shall show that there are parameter regions where the singlet-doublet decoupling holds and the realistic vacuum is a global minimum. Despite the
special nature
%
%
of the limit where our finding applies, this constitutes a proof of existence of a vacuum that is both stable and able to accommodate  $m_h\approx125\rm GeV$.
The results may be generalized to other cases as well.

Notice that in order for our approach to be self-consistent, {\em other} extrema, i.e., including $(s=\sfone, v=\vfone)$, also needs to satisfy the decoupling condition $\mathcal{M}^2_{13}=0$, just like the realistic minimum:
\beq
 \veff\left[2\mueff-\sintwobetaeff (A_\lambda+\frac{2\kappa}{\lambda}\mueff)\right]=0\label{m130}\,.
\eeq
This more restrictive condition of vanishing mixing entry also disfavors tachyons at other extrema which makes them more ``competitive'' compared to realistic vacuum. Meanwhile such additional condition constrains input parameters $A_\lambda, A_\kappa$ to satisfy certain relations with $\lambda, \kappa$ and output quantities at realistic vacuum.

In this limit  the stationary point equation of $S$ eq.~(\ref{minisinglet}) reduces to
 \beq
    s(2\kappa^2s^2+ A_\kappa s+m_S^2)=0\,,
 \eeq
and the solutions $\lbrace 0, \sr, s_{\textrm{u}} \rbrace$  take a simple form. Evidently $s_\textrm{u}$ satisfies:
\beq
s_{\textrm{u}}+\sr = -A_\kappa/(2\kappa)\label{quardsum}\,.
\eeq
We solve for nonzero $s_{\textrm{u}}$ using eqs.~(\ref{sin2beff},\ref{mzeff},\ref{m130}). First, $A_\lambda$ is fixed by the decoupling relation eq.~(\ref{m130}) and  $\mur, \sintwobetar, \lambda, \kappa$. Then based on eq.~(\ref{mzeff}) and eq.~(\ref{m130}) we can rewrite $\veff$ and $\mueff$ as function of $\sintwobetaeff$, then plug in eq.~(\ref{sin2beff}). A consistency equation for $\sintwobetaeff$ only can be obtained. We solve this equation
 and pick out 3 real solutions: $0, \sintwobetar, \sin2\beta_{\textrm{u}}$. The corresponding solutions for $(\seff, \veff)$ can be easily derived.

In order to compute the potentials we need to retrieve the remaining model parameters $A_\kappa, m_S^2$. We plug $s_\textrm{u}$ into eq.~(\ref{quardsum}) to find the model parameter $A_\kappa$. Plugging in $\mzr, \sintwobetar, \sr$ at realistic vacuum and $\lambda, ~\kappa, ~A_\lambda, ~A_\kappa $ into eq.~(\ref{ms2}), we retrieve the last model parameter $m_S^2$. Now we have obtained all 5 other sets of stationary solutions $(s, v)$ and can compute potentials using eq.~(\ref{potential}) and see when the realistic vacuum has the deepest potential, i.e., it is the global minimum.

\begin{figure}[t]
\begin{center}
\includegraphics[width=0.47 \linewidth]{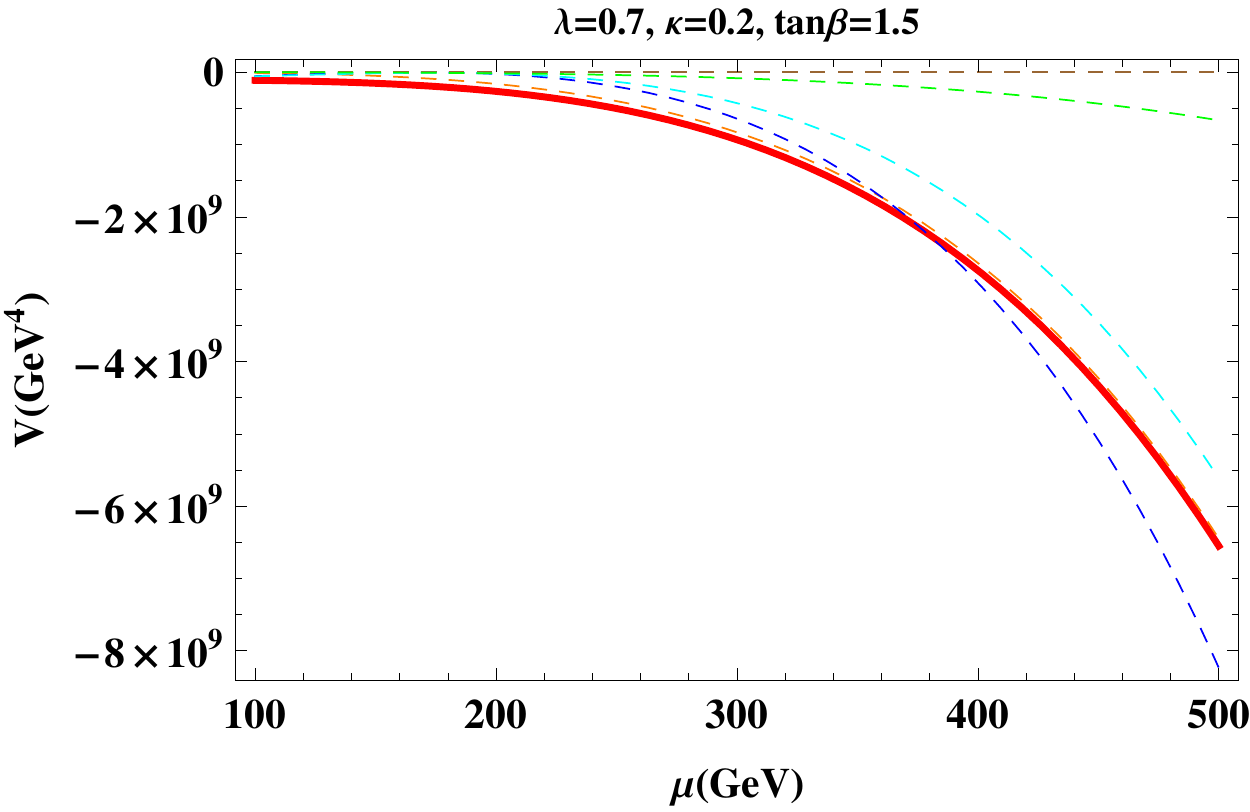}\hfill\includegraphics[width=0.49 \linewidth]{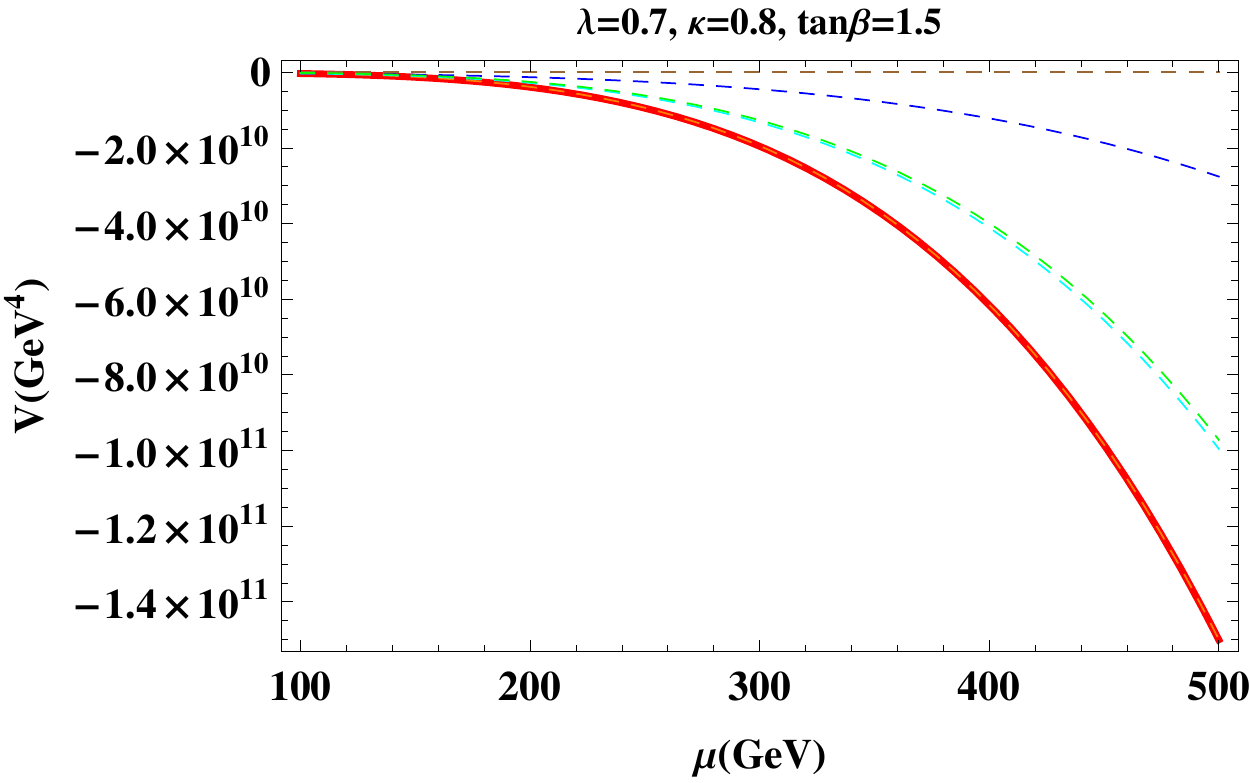}
\caption{The value of the potential evaluated at all possible stationary points defined by the requirement of having $\mu_{r}=\mu$ at the realistic minimum. The parameters of the potential are fixed as functions of $\mu_{r}$ as discussed in the text. In red and thick line we show the realistic minimum $(s=\sr, v=\vr)$. The other extremal points of the potential are detonated by dashed line with the following color code:  orange for $(s=\sr, v=0)$;  blue for  $(s=0, v=\vfzero)$;  brown for the origin $(s=0, v=0)$; cyan for $(s=\sfone, v=\vfone)$; green for $(s=\sfone, v=0)$. The orange line is close to red line in particular at large $\kappa$ as a result of the fact that the potentials of the corresponding extrema only differs by $\mzeff$ dependent terms in the potential eq.~(\ref{potential}) which is a sub-leading contribution. }
\label{globalminimum}
\end{center}
\end{figure}

The result of our analytic study  is shown in Figure~\ref{globalminimum}, where we evaluate the value of the potential at all the stationary points as a function of $\mur$, i.e., with all the values of the input parameters $A_\lambda, A_\kappa, m_S^2$ fixed as functions of $\mur, \tan\beta, \kappa, \lambda$ as explained above.
The red line
 corresponds to the value of the potential at the realistic realistic minimum. For different choices of $\kappa$ in the two panels this figure illustrates for what range of $\mur$ the realistic minimum is the global one.

In the figure we observe the danger of taking large $\mur$ discussed in Section \ref{s=0} above. The left panel shows that the value of the potential at the extremal point $(s=0,v=\vfzero)$ gets deeper than the realistic minimum when $\mur$ is large.
The trend as we increase $\kappa$ is as expected
from the earlier discussion: the realistic minimum is shallower
at smaller $\kappa$, and larger $\kappa$
is found to disfavor the unrealistic extremal point $(s=0,v=\vfzero)$.

\section{Push-up Scenario \label{pushupsection}}

As mentioned, one possibility of alleviating the tension between perturbativity up to GUT scale and $m_h\approx125\rm ~GeV$ is the ``push-up'' scenario.
As discussed earlier, $H_v^0$ is more decoupled from $h^0_v$ and $h_s^0$, in particular at small $\tan\beta$
as favored in the NMSSM to obtain large enough Higgs mass.
Therefore we focus on the 2-by-2 mass matrix in the basis of $(h^0_v, h_s^0)$. In this region of parameter space the SM Higgs-like state (mostly $h^0_v$) is typically the second lightest state of the spectrum~\footnote{We find that in the alternative case where the SM Higgs-like state is the heaviest state of the spectrum, the charged Higgs would be too light to be consistent with direct searches and flavor constraints such as from $b\rightarrow s\gamma$.}, {\em heavier} than the singlet-like state (mostly $h^0_s$), thus the mixing with the lightest state
induces an increase in its mass.
Thus, we must have
\begin{equation}
\mathcal{M}^2_{11} > \mathcal{M}^2_{33} \label{push-updef} \,.
\end{equation}
Apparently, it seems plausible to avoid the tuning associated with a small $\delta m^2_{ h \; \rm mix}$ in this case.
However, as we shall demonstrate in this section,
this scenario is strongly constrained by the LEP bound. In the end, the allowed push-up effect is very limited which corresponds to fine-tuning worse than  approximately 20\%~\footnote{See Ref.~\cite{Jeong:2012fk} for a similar analysis.}.

In detail, we have the following sub-mass matrix for $\tan \beta \approx 1$:
\beq
 \left(
   \begin{array}{cc}
     \mathcal{M}^{2}_{11} & \mathcal{M}^{2}_{13}  \\
     \mathcal{M}^{2}_{31}  & \mathcal{M}^{2}_{33}  \\
   \end{array}
 \right)
 =\left(
                 \begin{array}{cc}
                   \lambda^2v^2 & 2\lambda^2sv-(2\lambda\kappa s v+\lambda A_\lambda v) \\
                   \cdot & 4\kappa^2s^2+A_\kappa\kappa s+\frac{v^2}{2s}A_\lambda\lambda \\
                 \end{array}
               \right)\label{scalarmatrix22} \,.
\eeq
This tree-level mass matrix has two eigenvalues, $m_{1}$ and $m_{2}$, that are the masses of the doublet-like and singlet-like state, respectively. In this case $m_{1}\simeq m_{h}$. Also, in the push-up scenario $m_{1}>m_{2}$.

 Loop corrections to the masses from the stop sector are known to be substantial.  As mentioned earlier, for a representative analytic study we choose $m_{\tilde{t}}\sim500\rm ~GeV$ and negligible $A_{t}$. Such mass of the stop  gives a substantial correction to the Higgs mass yet still being compatible with naturalness.  Including the loop correction from such stop, we can aim at the tree-level Higgs mass
\begin{equation}
m_1\approx110 \textrm{ GeV}\,. \label{shootfor}
\end{equation}

In the push-up scenario the mass of the doublet-like Higgs is increased by mixing effects, therefore the range of the coupling $\lambda$ of interest is given by
 \begin{equation}
   \lambda v < m_{1} \approx 110 \textrm{ GeV} \label{apush-up1}
 \end{equation}
and thus
 \begin{equation}
\lambda \lesssim 0.7 \,.
\label{lambdalimit}
\end{equation}
 Also, to achieve the required spectrum with a lighter singlet-like state, we require the condition of eq.~(\ref{push-updef}) which explicitly reads as
 \begin{equation}
   \lambda^2v^2>4\kappa^2s^2+A_\kappa\kappa s+\frac{v^2}{2s}A_\lambda\lambda\label{push-upcondition} \,.
   \end{equation}

\bigskip

The lightest  CP-even Higgs of the push-up scenario is typically in the kinematical reach of LEP since  $m_2<110\rm~ GeV$.  This state is in general a mixture of singlet and doublet, due to the latter component it is constrained by LEP limits reported in  Figure~2 of Ref.~\cite{Schael:2006cr}. As we will see, a significant portion of the  parameter space of the push-up region is excluded by LEP.

To get an analytic understanding, we approximate the upper limit from LEP on the singlet-doublet
mixing
from above reference as a step-function of the lighter scalar mass:
\bea
 \textrm{Region-I}   &&\sin^2\theta
 \approx
  ~~0.01, \quad\quad 0<m_2<80 \gev \quad   \,,  \nonumber \\
 \textrm{Region-II}    &&\sin^2\theta
 \approx
  ~~0.1, \quad\quad 80 \gev <m_2<100 \gev \quad   \,,    \label{lepbound} \\
  \textrm{Region-III}  &&\sin^2\theta 
  \approx
  ~~0.4, \quad\quad 100 \gev <m_2<110 \gev \quad \,,  \nonumber
\eea
where $\theta$ is the mixing angle between the singlet and doublet in the matrix eq.~(\ref{scalarmatrix22}).
Note that (for all values of $\theta$) we have
$\tan2\theta=2\mathcal{M}^{2}_{13}/(\mathcal{M}^{2}_{11}-\mathcal{M}^{2}_{33})$.
Thus, the bound on mixing angle can be phrased directly as a constraint on the matrix elements in eq.~(\ref{scalarmatrix22}).
In turn, the constraint on mass matrix elements can be  mapped onto constraints on the model parameters such as
$\lambda$, $\kappa$ etc.
We can then get an idea of the tuning level
using the same parametrization discussed in Section \ref{sec:model}.

The
quantification of this tuning is the main goal of this section.
For this purpose  we shall compute numerically $\Delta_{\rm mix}$ as defined in eq.~(\ref{tuning}).
However in order to facilitate the analytic understanding, it is convenient to define a simplified fine-tuning formula in terms of mass matrix elements. We start from the reduced 2-by-2 mass matrix eq.~(\ref{scalarmatrix22}). In case of Region-I and II we may expand mass eigenvalues w.r.t. small mixing angle. Furthermore to obtain analytically tractable expressions we need to assume $\mathcal{M}^{2}_{33} \ll \mathcal{M}_{11}^{2}$. The latter assumption is not necessarily true in all the parameter space, however it can be realized in large fraction of the relevant  parameter space since the push-up condition requires $\mathcal{M}^{2}_{11}>\mathcal{M}^{2}_{33}$. With $s\approx v$~\footnote{From chargino bound on $\mu$ term, we know $s\gtrsim v$, so taking $s\approx v$ gives a \textit{conservative} estimate of the level of tuning here. Furthermore, as we will see from numerical analysis section, after including other constraints such as no-tachyons, $\mu$ term has to be $\lesssim 120 \rm GeV$ which means $s\approx v$.} we get the approximate expression derived from eq.~(\ref{tuning}):
 \beq
 \Delta_{\rm mix}\simeq\frac{4\mathcal{M}^{2}_{11}}{\mathcal{M}^{2}_{13}}\,. \label{simpletuning}
 \eeq
 For later convenience we also rewrite this fine-tuning as
 \beq
  \Delta_{\rm mix}\simeq\frac{4\mathcal{M}^{2}_{11}\theta}{\mathcal{M}^{2}_{11}-m_1^2}\,.
  \label{tuningestimate}
 \eeq
It is easy to check that this closely reproduces the $\Deltamix$ that we would obtain from the standard definition eq.~(\ref{tuning}).

\bigskip

We first consider the small mixing region including Region I and II in eq.~(\ref{lepbound})
where $\sin^2\theta\lesssim0.1$ and $m_2<100\rm ~GeV$.
We then have $\tan 2 \theta
\ll1$ so that
$\theta\approx \mathcal{M}^{2}_{13}/(\mathcal{M}^{2}_{11}-\mathcal{M}^{2}_{33})$.
We can expand the mass eigenvalues in a series in
$\theta$:
\bea
m_1^2&=&\mathcal{M}^{2}_{11}+\theta^2(\mathcal{M}^{2}_{11}-\mathcal{M}^{2}_{33})\label{m1mix01} \, ,\\
m_2^2&=&\mathcal{M}^{2}_{33}-\theta^2(\mathcal{M}^{2}_{11}-\mathcal{M}^{2}_{33})\label{m2mix01}\,.
\eea
From eq.~(\ref{m1mix01}) we get
\beq
\mathcal{M}^{2}_{33}=\frac{(1+\theta^2)\mathcal{M}^{2}_{11}-m_1^2}{\theta^2}\,
\eeq
and plugging this back into eq.~(\ref{m2mix01}) we obtain
 \beq
 m_2^2=\frac{1+2\theta^2}{\theta^2}\mathcal{M}^{2}_{11}-\frac{1+\theta^2}{\theta^2}m_1^2\label{m2master} \,.
 \eeq

From this we get bounds on $\mathcal{M}^{2}_{11}$ and on the degree of tuning at the boundary (i.e.,
maximum value of $\theta$) of different mass regions~\footnote{Tuning is clearly worse for smaller values of
$\theta$.}:
\bit
\item Region-I: \quad $\sin^2\theta=0.01, 0<m_2<80 \gev , m_1=110\gev$\\
  Plugging this in eq.~(\ref{m2master}) we get

\begin{eqnarray}
  109.5^2\gev^{2} & \lesssim \mathcal{M}^{2}_{11}\lesssim & 109.7^2\gev^{2}\,,\nonumber \\
  24^2 \gev^{2} & \lesssim \mathcal{M}^{2}_{13}\lesssim &34^2 \gev^{2}\,, \nonumber\\
   11^2 \gev^{2} & \lesssim \mathcal{M}^{2}_{33}\lesssim &80^2 \gev^{2}\,,\label{regionI}
  \end{eqnarray}
that corresponds to at best about 2\% tuning (i.e., $\Deltamix\approx40$) from eq.~(\ref{simpletuning}).

\item Region-II: \quad $\sin^2\theta=0.1, 80 \gev <m_2<100 \gev, m_1=110 \gev$\\
  Plugging in eq.~(\ref{m2master}) we get

\begin{eqnarray}
  108^2 \gev^{2} & \lesssim \mathcal{M}^{2}_{11}\lesssim & 109^2 \gev^{2}\,, \nonumber \\
   24^2 \gev^{2}  & \lesssim \mathcal{M}^{2}_{13}\lesssim & 39^2 \gev^{2} \,,\nonumber\\
   83^2 \gev^{2} & \lesssim \mathcal{M}^{2}_{33}\lesssim &101^2 \gev^{2}\,. \label{regionII}
  \end{eqnarray}
This would correspond to a tuning of about 5\%~($\Deltamix=20$). However we notice that the obtained ranges for the matrix elements are incompatible with the assumptions that we made to get an analytically manageable expression for the tuning. Therefore this estimate of the tuning in Region-II is expected to be very crude. In fact the numerical results given later show that the tuning can be as good as 20\% ($\Deltamix=5$).

\eit

Here a few comments are in order.  From the above analysis $\mathcal{M}^{2}_{11}$ can only lie in the narrow region $108^2-110^2$ which directly tells $\lambda\approx0.6$, and not smaller. Additionally we remark that the closeness of $\mathcal{M}^{2}_{11}$ to $m_1$ and consequently the tuning is a result of the constraint  from LEP. 

These bounds on the mass matrix entries, covering regions I {\em and} II can be translated into  constraints on the model parameters $\lambda, \kappa, A_\kappa, A_\lambda$. The discussion is divided into several sub-cases depending on which {$A$-terms} are vanishing. This division is general from a bottom-up point of view and also can be easily mapped onto complete models of SUSY breaking where a typical size of the $A$-terms is predicted.

 \bit
 \item Case-I:  $A_\kappa=0, A_\lambda=0$ \\
       The small value(s) of the bound on $\mathcal{M}^{2}_{13}$ in  eq.~(\ref{regionI}) and  eq.~(\ref{regionII})
      combined with the expression for it in     eq.~(\ref{scalarmatrix22})
      and $\lambda\approx0.6$ as derived above implies
       $$\lambda\approx\kappa\approx0.6\,.$$ We know from chargino LEP bound that $\mu=\lambda s\gtrsim 105\rm~GeV$, thus $s\gtrsim v$. Meanwhile $\mathcal{M}^{2}_{33}=4\kappa^2s^2$, hence $\lambda\approx\kappa$ implies $\mathcal{M}^{2}_{33}>\mathcal{M}^{2}_{11}$ contradicting both
       the push-up condition $\mathcal{M}^{2}_{11}>\mathcal{M}^{2}_{33}$ given in eq.~(\ref{push-upcondition}) and the $\mathcal{M}^{2}_{33}$ bound that we just derived as in eq.~(\ref{regionI}) or eq.~(\ref{regionII}).
       In conclusion this region does not work in any way.
 \item Case-II: $A_\kappa\neq0, A_\lambda=0$\\
       In this case we also need $$\lambda\approx\kappa\approx 0.6$$
       since $A_{ \kappa }$ does not enter $\mathcal{M}^{2}_{13}$.
       The push-up condition and $\mathcal{M}^{2}_{33}$ bound can now be accommodated by a suitable choice of $A_\kappa$. However,
       we remark that this case is not viable if we require perturbativity up to GUT scale because  $\kappa$ is too large.
 \item Case-III: $A_\kappa=0, A_\lambda\neq0$\\
In this case the small $\mathcal{M}^{2}_{13}$ is attained requiring $(\lambda-\kappa)s-A_\lambda/2\approx 0$. When  $\kappa$ is taken small the model may be perturbative up to the GUT scale. In this case a tuning between $\mu$ and $A_\lambda$ is required. Later in our numerical studies we will find this tuning to be 20\% or worse ($\Deltamix \geq 5$). The chosen $A_\lambda$ also needs to satisfy the bound on $\mathcal{M}^2_{33}$ as derived in eq.~(\ref{regionI}) or eq.~(\ref{regionII}) as well as the push-up condition $\mathcal{M}^2_{11}>\mathcal{M}^2_{33}$.\\

Of course we would expect the combination of Case-II and Case-III ($A_\kappa\neq0, A_\lambda\neq0$) may contain viable parameter space as well. But we leave this to our numerical study as the analytic study gets complicated in this combined case.

 \eit

\bigskip

For Region-III the mixing is large, $\tan2\theta>1$ ($\sin^2\theta\gtrsim0.15$). In this case the proper way to get simplified mass eigenvalues is to expand in a series in $\epsilon \equiv (\mathcal{M}^2_{11}-\mathcal{M}^2_{33})/(2\mathcal{M}^{2}_{13})<1$:
\bea
  m_1^2
   &=&\frac{1}{2}\left[\mathcal{M}^2_{11}+\mathcal{M}^2_{33}+2\mathcal{M}^2_{13}\left(1+\frac{(\mathcal{M}^2_{11}-\mathcal{M}^2_{33})^2}{4(\mathcal{M}^2_{13})^2}\right)\right]  \nonumber \\
   &=&\frac{1}{2}\left[(1+\frac{1}{\epsilon})\mathcal{M}^2_{11}+(1-\epsilon-\frac{1}{\epsilon})\mathcal{M}^2_{33}\right]\label{m1mix2} \,,\\
  m_2^2
    &=&\frac{1}{2}\left[\mathcal{M}^2_{11}+\mathcal{M}^2_{33}-2\mathcal{M}^2_{13}\left(1+\frac{(\mathcal{M}^2_{11}-\mathcal{M}^2_{33})^2}{4(\mathcal{M}^2_{13})^2}\right)\right]  \nonumber \\
    &=&\frac{1}{2}\left[(1-\frac{1}{\epsilon})\mathcal{M}^2_{11}+(1+\epsilon+\frac{1}{\epsilon})\mathcal{M}^2_{33}\right] \,. \label{m2mix2}
\eea
Eq.~(\ref{m1mix2}) gives
\beq
\mathcal{M}^2_{33}=\frac{2m_1^2-(1+\frac{1}{\epsilon})\mathcal{M}^2_{11}}{1-\epsilon-\frac{1}{\epsilon}} \,.
\eeq

In this region, expanding the eigenvalues of 2-by-2 mass matrix w.r.t $\epsilon$ and further simplify with $\mathcal{M}^2_{11}>\mathcal{M}^2_{33}$, we derived an approximate tuning formula from eq.~(\ref{tuning}):

\beq
 \Delta_{\rm mix}\equiv\frac{2\mathcal{M}^2_{11}}{\mathcal{M}^2_{13}}\,. \label{simpletuninglargemixing}
 \eeq

Plugging this  into eq.~({\ref{m2mix2}})  and applying $\epsilon=1/5$ ($\sin^2\theta\approx0.4$), and $100\gev \lesssim m_2\lesssim110\gev$, we get the following bounds on $\mathcal{M}^2_{11},\, \mathcal{M}^2_{33}$:
\bit
\item Region-III: \quad $\sin^2\theta\approx0.4, 100\gev\lesssim m_2\lesssim110\gev, m_1=110 \gev $

  \begin{eqnarray}
  107^2 \gev^{2} & \lesssim   \mathcal{M}^2_{11}\lesssim  & 111^2 \gev^{2}\,,\nonumber \\
  34^2\gev^{2}    &  \lesssim  \mathcal{M}^2_{13}\lesssim  & 46^{2} \gev^{2}\,, \nonumber  \\
  103^2 \gev^{2} & \lesssim   \mathcal{M}^2_{33}\lesssim  &109^{2} \gev^{2}\,,
  \end{eqnarray}
that  corresponds to a tuning at best about $10\%$ (i.e., $\Deltamix\approx10$)
from eq.~(\ref{simpletuninglargemixing}).
Notice that now $\mathcal{M}^2_{11}$ and $\mathcal{M}^2_{33}$ have to be near degenerate, which might be seen as  another kind of tuning.
\eit

Some comments on the above large mixing case are in order.
Despite of the large mixing, the bound on $\mathcal{M}^2_{11}$ tells that we still need $$0.6\lesssim\lambda\lesssim 0.65\,.$$ and thus the push-up effect is limited. This is the case because the large mixing mostly can only be due to the near degeneracy of the masses of the two eigenstates in order to avoid LEP limits. Consequently, although the fine-tuning defined by $\Delta_{\rm mix}$ is slightly better than the small mixing case of Region-I, it is still at the level of $10\%$  or worse.
As far as the discussion of tuning is concerned, it is worth stressing that, while the larger mixing slightly alleviate the fine-tuning of  $\delta m^2_{ h \; \rm mix}$ of eq.~(\ref{schematic})
at the same time it requires another  kind of tuning such that $\mathcal{M}^2_{11}$ and $\mathcal{M}^2_{33}$ are near degenerate as mentioned earlier.
A simple estimator of the precision of the cancellation required here may be $(\mathcal{M}^2_{11}-\mathcal{M}^2_{33})/(\mathcal{M}^2_{11}+\mathcal{M}^2_{33})$~--~which is found to be about $5\%$ tuning or worse.

It is also handy to obtain constraints on other model parameters in this large mixing case.
When the $A$-terms are zero, to get the mixing angle one needs $0.55\lesssim\kappa\lesssim0.6$, which is still very close to $\lambda$ and violates the push-up condition eq.~(\ref{push-upcondition}), analogous to the case of small mixing, and thus is not viable at all.
For non vanishing $A$-terms the discussion also mostly follows the one presented for the small mixing case of Region-I and Region-II.

 \bigskip

 The analytic considerations exposed above are powerful estimators of the degree of tuning necessary to obtain a viable spectrum with the lightest CP-even state
 being sufficiently singlet-like to avoid LEP limits and a heavier state with a mass of 125 GeV, i.e., the SM-like Higgs boson
 observed at the LHC.
 To verify our findings and to cover the intermediate cases that do not clearly belong to one of the limits studied above,
 we next perform a {\em numerical} study of the properties of the NMSSM in the region of the  parameter space that
 includes
 the cases outlined above.

The study is performed {\em fixing} the parameters  \begin{center}$\mu$, $A_{\lambda}$, $A_{k}$, $\tan\beta,$\end{center} while we let  free the parameters $\lambda$ and $\kappa$. All the observables are computed with no approximations.

Following the discussion on the parametric behavior of the Higgs masses and mixing presented above we discuss the following points in the parameter space, in order to illustrate the cases with   $A_{\lambda}\neq 0,\, A_{k}\ll A_{\lambda}$, with $A_{\lambda} \ll A_{\kappa},\, A_{k}\neq0$, and with generic $A$-terms, respectively:
\bit
\item $\mu=120 \gev,\, A_{\lambda}=250\gev,\, A_{k}=0\,, \tan\beta=1.5 $
\item $\mu=110 \gev,\, A_{\lambda}=50\gev,\, A_{k}=-250\gev\,, \tan\beta=1.5$
\item $\mu=120 \gev,\, A_{\lambda}=250\gev,\, A_{k}=-250\gev,\, \tan\beta=1.5$
\eit

\begin{figure}[h!]
\begin{centering}
\includegraphics[width=0.45\linewidth]{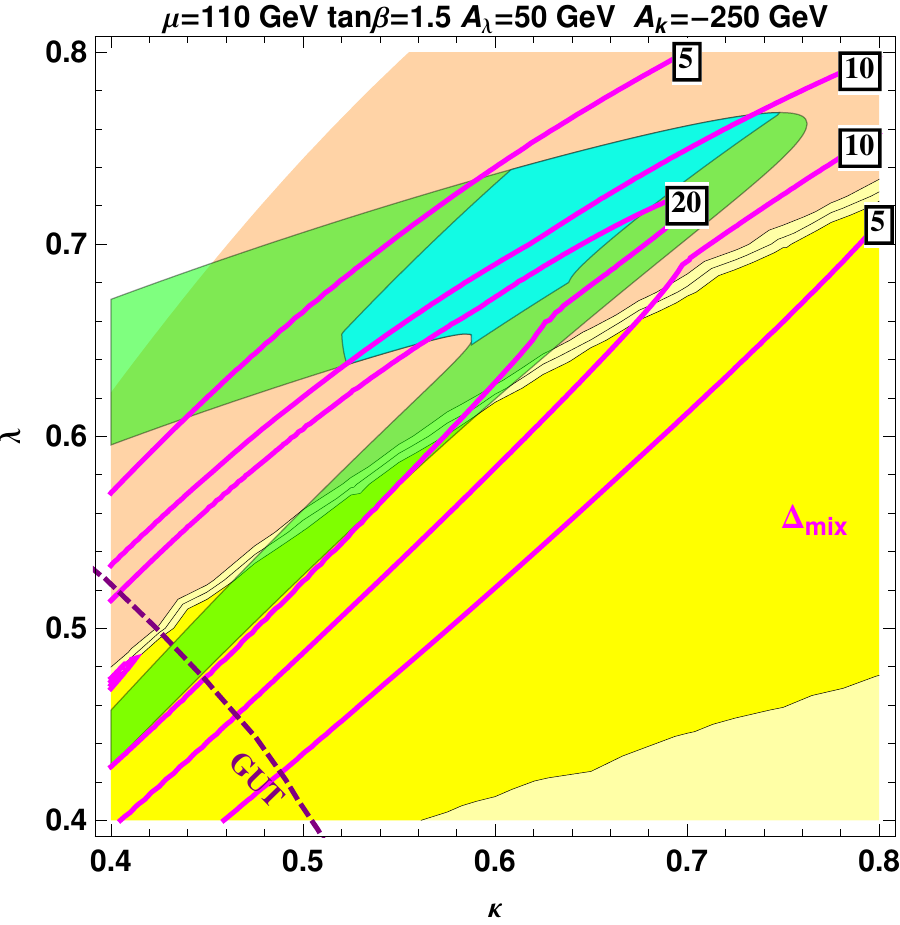}\includegraphics[width=0.45\linewidth]{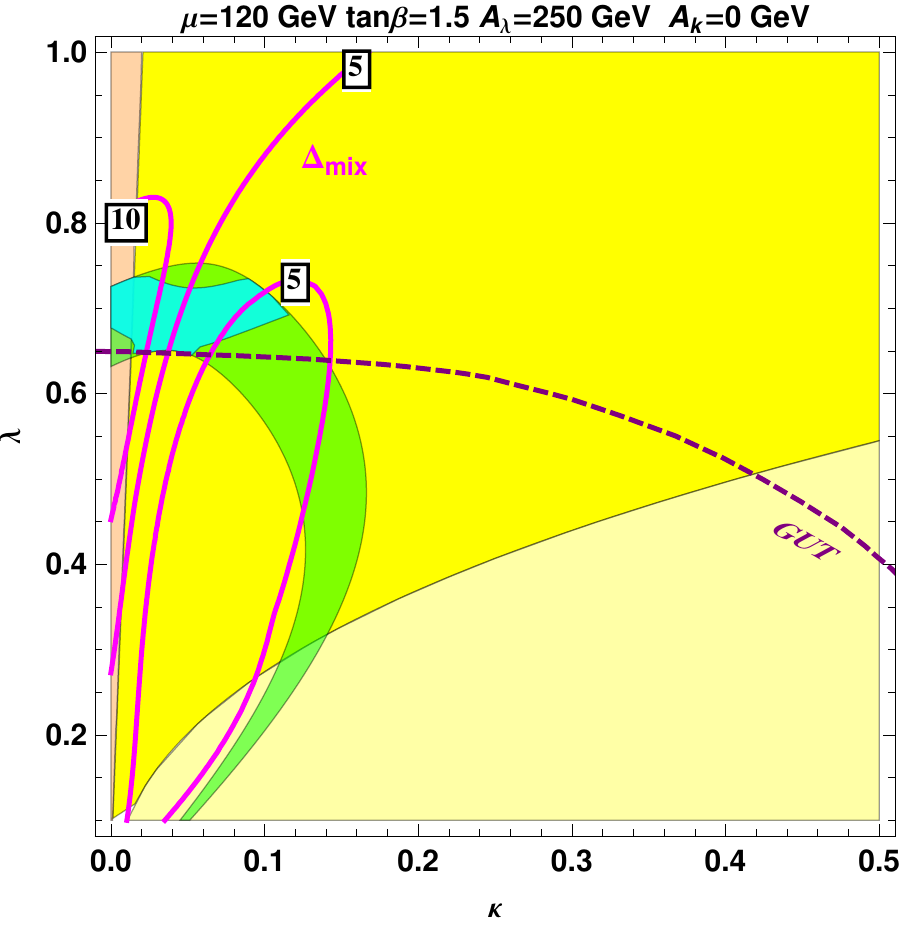}
\par\end{centering}

\begin{centering}
\includegraphics[width=0.45\linewidth]{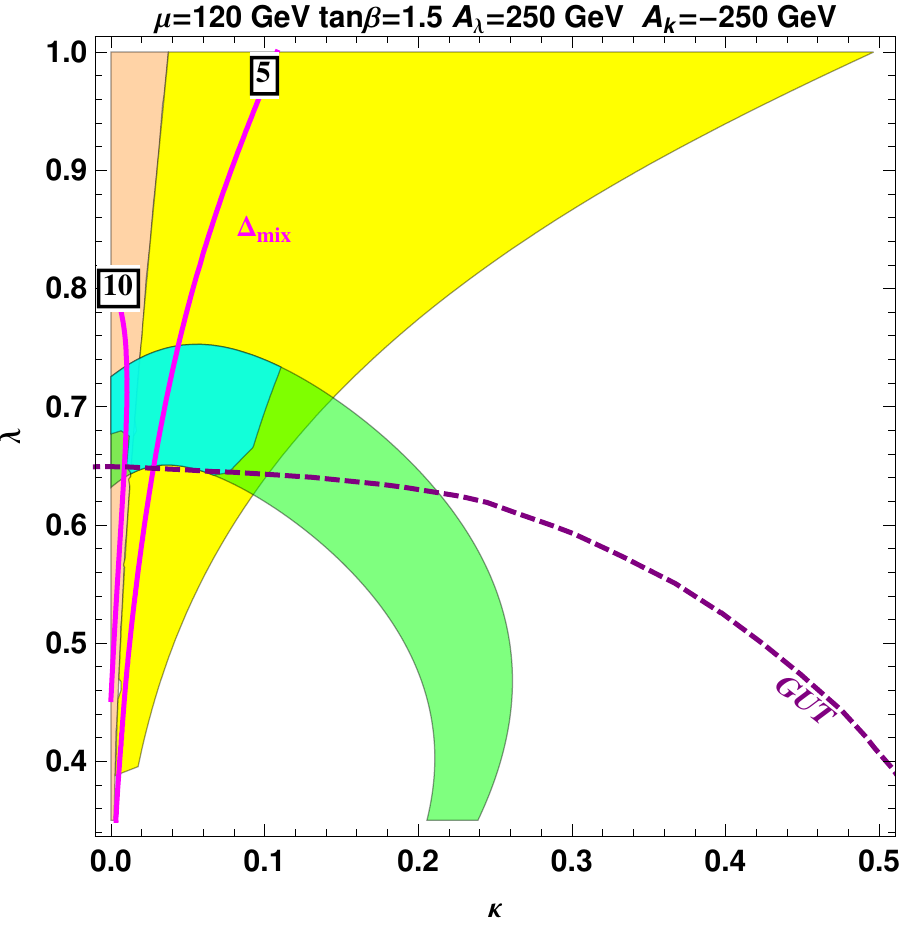}
\par\end{centering}

\caption{\label{plotspushup} Constraints and fine tuning in the $\left(\lambda,\, \kappa\right)$ plane in the push-up
scenario. The solid magenta lines are the isolines of the fine tuning
measure $\Deltamix$ defined in eq.~(\ref{tuning}). The cyan region is the region compatible with both LEP searches for the lightest CP-even and with the tree-level mass of the SM-like state in the range 110~-~125 GeV as suggested by LHC data. The green region would be compatible with LHC but is ruled out by LEP
 (based on the bounds in eq.~(\ref{lepbound})). 
 \protect \\
The light and bright yellow and the orange areas denote the regions where the realistic minimum is an acceptable
vacuum according to the requirements of eq.~(\ref{eq:StandardVacuumConditions}).
In the orange region other minima where $v\neq \vr$ or $s\neq \sr$
(or both) are significantly (more than 5\%) deeper than the realistic minimum. In the light yellow region the extra-minima
are degenerate within 5\% with the realistic minimum. In the bright yellow
region the extra minima are either absent or significantly shallower
than the realistic minimum.
The purple dashed line is the boundary of the region that contains
couplings that stay perturbative (less than $\sqrt{2\pi}$) up to
$\Lambda_{GUT}=1.6\cdot10^{16}$ GeV.
}
\end{figure}

We shall  identify the region in the $(\lambda\,,\kappa)$ plane that is compatible with a number of requirements of the scalar potential. In particular we shall require that the the point $(H_{u}\,,\, H_{d})=(0,0)$ in the fields space is unstable, in order to favor EW symmetry breaking. For the same reason we shall require the potential to have a minimum at the point where the singlet field scalar field is $\langle S \rangle=s\equiv{\mu\over\lambda}$ and the doublet fields are at the point $(\langle H_{d}\rangle\,,\,\langle H_{u}\rangle)=(v\sin\beta\,,\,v\cos\beta)$, where $v\simeq 174 \gev$.
At the realistic minimum we must also make sure that  the $U(1)$ of the electromagnetism is not broken. All in all we require the following basic conditions for a realistic potential:
\begin{eqnarray}
& (H_{u},\, H_{d})=(0,\,0)\textrm{ is an unstable point}    \nonumber  \\
& V(v,s)<0   \nonumber \\
& m_{s_{1}}^{2}>0,\, m_{a_{1}}^{2}>0,\, m_{H^{\pm}}^{2}>0  \, .\label{eq:StandardVacuumConditions}
\end{eqnarray}
In Figure~\ref{plotspushup} we show in bright yellow, light yellow and orange regions where all these conditions are satisfied.
The difference between these three regions lies in the additional requirements on the stability of the realistic minimum.
Indeed, the potential of the NMSSM can have several minima and, as we discussed in Section \ref{extraminima},  the realistic minimum  may not be the global minimum, posing a question on the stability of the vacuum where we are supposed to live. The points in the orange region are points at which the potential has minima deeper than the realistic one, i.e., the realistic minimum is a local but not a global minimum. In the region in bright yellow the realistic minimum is the global minimum of the potential while in the light yellow region the realistic minimum is almost degenerate with another minimum. In particular in the light yellow region the two minima (regardless of which is the deeper) are degenerate up to a 5\% difference in their depths, i.e.,
\begin{equation}
|V_{1}/V_{2} -1| <0.05 \label{treelevelunreliable}\,,
\end{equation}
where $V_{1,2}$ are the depth of the two minima. This 5\% threshold is chosen as we expect that loop corrections will
change the depths of the minima by a few percent and therefore it is only when eq.~(\ref{treelevelunreliable}) does not hold that we can use the tree-level potential to reliably determine what minimum is the global one.

To firmly exclude or allow some of these points, a proper analysis would be required, namely, the computation of the lifetime of the vacua to establish if we are living in a vacuum that is sufficiently long-lived~\footnote{
Even that would be just a necessary check for the viability of these points, since other pathologies may arise. For instance, there is still the risk that in the early Universe the scalar fields stay at the unrealistic minimum for too  long  and significantly alter the many successful predictions of standard cosmology.}.
 The lifetime can be estimated using the Euclidian action of the field configuration that interpolates between the two minima, a task that we leave for future work. However it is clear that the points in the orange region are likely to have too a short lifetime for the realistic vacuum and are in some sense disfavored.

In Figure~\ref{plotspushup}  we show in cyan the region allowed by LEP and LHC data on the SM-like Higgs state and singlet-like state in the push-up scenario. We also show the isolines of the fine-tuning computed with the exact formula eq.~(\ref{tuning}). The purple dashed line is the boundary of the region that contains
couplings that stay perturbative (less than $\sqrt{2\pi}$) up to
$\Lambda_{GUT}=1.6\cdot10^{16}$ GeV according to a one-loop RGE with
boundary conditions at 1 TeV and 4 sets of $5+\bar{5}$ of SU(5) entering
in the RGE at 5 TeV.

Figure~\ref{plotspushup} shows that, according to our estimates above, in the push-up scenario it is possible to have a lightest singlet-like CP-even scalar compatible with LEP and a heavier SM Higgs-like state at the mass suggested by the recent discovery of the LHC experiments.
The cases where $A_{\lambda}$ is chosen to obtain a small $\mathcal{M}^2_{13}$ are the least tuned, i.e., there are
phenomenologically viable points of this type having $\Deltamix \gtrsim 5$.
Furthermore for sizable $A$-terms it becomes possible for the model to be perturbative up to the GUT scale. 
However we remark that the region compatible with perturbative unification is very narrow. Therefore we conclude that there is a moderate tension between the requirement of perturbativity up to the GUT scale and the observed mass of the Higgs.
A caveat is in order about this result. The boundary of the perturbative region marked by a purple dashed line in the plot should be considered as a loose one, due to the inherent uncertainties in the definition of a Landau pole. For instance, changing the coupling at the Landau pole by just a factor 2 would result in a few percent difference in the coupling $\lambda$ at the TeV scale. 
However we do not expect these minor corrections to change our conclusion in a significant way.

Of course a detailed investigation of the second lightest CP-even scalar of the push-up scenario should be carried on to fully assess the compatibility of this region of parameter space with all the available information about the newly discovered SM-like Higgs
boson. We just remark that this state tends to be SM-like, as it cannot mix much with the singlet, and the other state doublet-like state is rather decoupled. Hence the 125 GeV boson in the push-up scenario seems in reasonable agreement with current observations of the properties of the new boson.
\newpage

\newpage

\section{Pull-down Scenario \label{pulldownsection}}

In the pull-down scenario the effect of the  mixing in eq.~(\ref{schematic}) is to reduce the mass of the SM-like Higgs boson
relative to the upper bound. As suggested by Figure~\ref{mhmax}, the observed mass of the Higgs then
requires $\lambda \gtrsim 0.7$.
We  divide the range of $\lambda$ into $\lambda \simeq 0.7$ and $\lambda > 0.7$. The reason to do this splitting is that the former value {\em can be} -- but is not guaranteed to be  -- perturbative up to the GUT scale, whereas the latter values are clearly non-perturbative.

We start with an analytic study in order to have a clear view before the numerics.
For this purpose, we assume a hierarchical mass squared matrix,  i.e.,
${\cal M}^2_{  11 } \ll {\cal M}^2 _{   22, 33 }$ and
${\cal M}^2_{  12 } \ll {\cal M}^2_ {  22}$, ${\cal M}^2_{  1 3 } \ll {\cal M}^2_{  33 }$.
Note that the above assumption are compatible with
 what is often referred to as ``decoupling limit'', i.e., $s \gg v$ with $\lambda \sim \kappa$,
as well as the case
 where all the states are mixed, $s \sim v$ with $\kappa \gg \lambda$.

As we have explained already we consider only the two state mixing problem for $h^0_v - H^0_v$. Hence the mass of the doublet-like state is
\begin{eqnarray}
m_h^2 & \approx & m_Z^2 \cos^2 2 \beta + \lambda^2 v^2 \sin^2 2 \beta \nonumber \\
& & - \lambda^2 v^2 \left( \sin 2 \beta - \frac{1}{ \rho } \right)^2\,,
\label{noA1}
\end{eqnarray}
where $\rho \equiv \kappa / \lambda$.
The first line above -- which is ${\cal M}^2_{11}$ from eq.~(\ref{CPeven}) -- is the usually quoted upper bound on SM-like Higgs mass in the NMSSM. The second line is the {\em negative} effect from mixing.

The above formula can be
generalized to include $A$-terms, assuming that the $A$-terms do not dominate over singlet VEVs:
\begin{eqnarray}
m_h^2 & \approx & m_Z^2 \cos^2 2 \beta + \lambda^2 v^2 \sin^2 2 \beta \nonumber \\
& & - \lambda^2 v^2 \frac{
\Big[
\sin 2
\beta
\left( 1 + \frac{ A_{ \lambda } }{ 2 \kappa s } \right) - \frac{1}{ \rho }
\Big]^2 }{ 1 +   \frac{ A_{ \kappa } } { 4 \kappa s } + \frac{ A_{ \lambda } \sin 2 \beta \; v^2 }{ 8 \kappa \rho s^3 } }
\label{withA}\,.
\end{eqnarray}

\subsection{$\lambda \simeq 0.7$}

In this case, ${\cal M}^2_{ 11 }$ is close to $( 110 \; \hbox{GeV} )^2$. Thus, we cannot afford a sizable pull-down effect so that the goal here is to {\em minimize} this effect and saturate the upper bound on SM-like Higgs mass. This minimization of the mixing is clearly a source of tuning, a \emph{tuning of the mixing} as we dub it in order
to distinguish it from other kind of tunings to be encountered later
and is quantified by $\Deltamix$ in eq.~(\ref{tuning}).
\bigskip

For negligible $A$-terms  we see from eq.~(\ref{noA1}), that the only option to minimize the pull-down effect is  to tune $\kappa$ vs. $\lambda$ such that
\begin{equation}
\kappa \approx \lambda / \sin 2 \beta\,. \label{lambdakappa}
\end{equation}
In this case, the value of $\kappa$
is
so large that the couplings become non-perturbative below the GUT scale.

\bigskip

For sizable $A$-terms, we can keep $\kappa$ small so that the theory is perturbative up to the GUT scale. Instead, we can tune $A_{ \lambda }$ vs. $\mu$ in order to
minimize the effect of mixing in eq.~(\ref{withA}):
\begin{equation}
A_{ \lambda } \approx 2 \mu / \sin 2 \beta\,.
\end{equation}
Another possibility would have been to make $A_{ \kappa }$ large and positive so as to suppress the mixing effect.
However for typical $A_{k}$ needed to reduce the mixing effect the CP-odd pseudo-scalar masses from eq.~(\ref{CPodd}) are  tachyonic, hence this option is not viable.

\subsection{$\lambda > 0.7$}

In this case the theory becomes non-perturbative below the GUT scale due to the large value of $\lambda$ at the EW scale, regardless of the value of $\kappa$.
The maximal Higgs mass attainable is now larger than 125 GeV, thus a certain amount of pull-down effect is
actually {\em needed}.
For this reason the \emph{tuning of the mixing} encountered for $\lambda \simeq 0.7$ is {\em absent} for larger $\lambda$.
therefore we begin the discussion with examples for $\lambda\simeq1$.

\bigskip

For negligible $A$-terms it is better to re-write the Higgs mass formula as
\begin{eqnarray}
m_h^2 & \approx & m_Z^2 \cos^2 2 \beta + \nonumber \\
& &  \lambda^2 v^2 \left( -\frac{1}{ \rho^2 } + \frac{2}{ \rho } \sin 2 \beta \right)\,.
\label{noA2}
\end{eqnarray}
Given the large value of $\lambda$, to obtain the desired mass of the Higgs we need the term in parenthesis to be much smaller than $1$ versus being $\approx1$ for $\lambda\lesssim0.7$.
For this purpose we need to deviate from eq.~(\ref{lambdakappa}). There are two options. The first one is to choose $\rho < 1 /\sin2\beta$ that is
\begin{equation}
\kappa < \lambda /  \sin 2 \beta  \,.
\end{equation}
The small $\kappa$ helps to keep the Landau pole far from the EW scale (even though still below GUT scale).
However, adjusting the parameters in this way tends to
require a cancellation between the two terms inside parenthesis in second line of eq.~(\ref{noA2}).
For example, $\lambda = 1$, $\tan \beta = 1.5$ requires $\kappa \simeq 0.65$ to get Higgs mass down from
upper bound of 164 GeV to 125 GeV which comes from a
cancellation of about 1 part in 5 between two terms in second line of eq.~(\ref{noA2}).
Note that this tuning is {\em different} from the tuning of the mixing considered earlier for the case with $\lambda \simeq 0.7$.

The second way to get the Higgs mass from eq.~(\ref{noA2}) is
to choose $\rho>1$, that is,
\begin{equation}
\kappa > \lambda /  \sin 2 \beta  \,.
\end{equation}
For this choice of parameters the tuning associated with cancellation in the second line
of eq.~(\ref{noA2}) is not an issue since the second term always dominates. However in this case
  $\kappa$ is larger than $\lambda$ and thus perturbativity is lost  rather close to the EW scale.
For example, $\lambda =1$, $\tan \beta = 1.5$ requires $\kappa \simeq 3.2$.
For examples of a possible UV completion of the NMSSM with choice of couplings  $\lambda \gtrsim 1$ and
$\kappa > \lambda$ see Refs.~\cite{Larsen:2012ys, Chang:2004db}.

Finally, in the range of $\lambda$ in between
0.7 and 1 there might be natural choices of parameters.
For example, $\tan\beta=1.5$, $\lambda = 0.77$, $\kappa = 0.67~\textrm{or}~1.1$, can get Higgs mass down from
upper bound of 129 GeV to 125 GeV which is associated with a
cancellation of about 1 part in 5.

\bigskip

\begin{figure}[h!]
\begin{centering}
\includegraphics[width=0.49\linewidth]{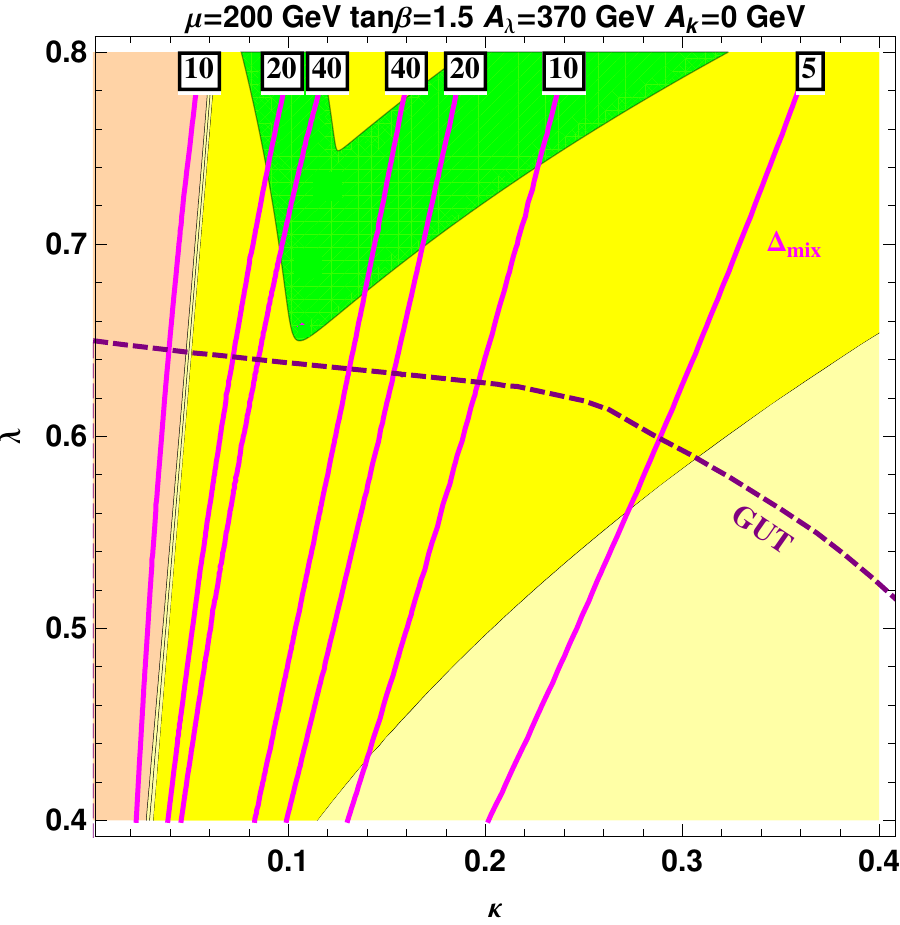} \hfill \includegraphics[width=0.49\linewidth]{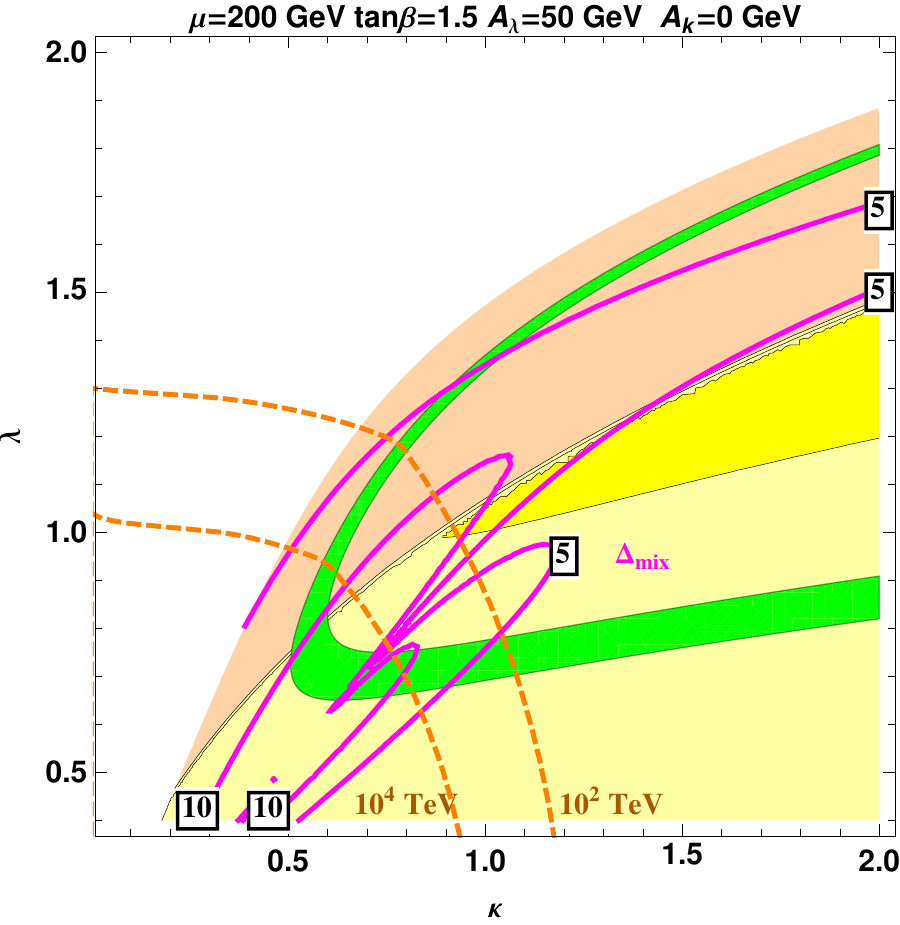}\par\end{centering}

\caption{
Constraints and fine tuning in the $\left(\lambda,\, \kappa\right)$ plane in the pull-down
scenario. The solid pink lines are the isolines of the fine tuning
measure $\Deltamix$ defined in eq.~(\ref{tuning}). The green region is the region where  the tree-level mass of the SM-like state in the range 110~-~125 GeV as suggested by LHC data and by Naturalness considerations on the mass of the stop.\protect \\
The light and bright yellow and the orange areas denote the regions where the desired minimum is an acceptable
vacuum according to the requirements of eq.~(\ref{eq:StandardVacuumConditions}).
In the orange region other minima where $v\neq \vr$ or $s\neq \sr$
(or both) are significantly (more than 5\%) deeper than the realistic minimum. In the light yellow region the extra-minima
are degenerate within 5\% with the realistic minimum. In the bright yellow
region the extra minima are either absent or significantly shallower
than the realistic minimum.
The purple dashed line is the boundary of the region that contains
couplings that stay perturbative (less than $\sqrt{2\pi}$) up to
$\Lambda_{GUT}=1.6\cdot10^{16}$ GeV.
The orange dashed lines in the right plot are the isolines of the
Landau pole scale $\Lambda_{LP}$ where one of the couplings gets non-perturbative.
\label{pulldownplots}
}
\end{figure}

Guided by the analytical intuition developed above we now proceed to the
numerical computation of several properties of the NMSSM.  As we did for the push-up scenario, in order to study the regions identified above we shall vary the parameters $\lambda$ and $\kappa$ while fixing the other parameters.
The right panel of Figure~\ref{pulldownplots} illustrates the case with {\em negligible} $A$-terms. Color codes for the global minimum analysis are the same as for the study of the  push-up scenario in Section \ref{pushupsection}. The green band in the plot is the region where the Higgs mass is in the range 110~-~125~GeV. As expected from our analytic study, in this case it is impossible to avoid a Landau pole below GUT scale.

 Indeed the Landau pole is rather low, as illustrated by the dashed lines that correspond to a Landau pole at $10^{4}$ or $10^{2}$ TeV.  If one imagines an embedding of the NMSSM in a model of supersymmetry breaking such as gauge mediation, where there is a meta-stable NLSP, there are interesting consequences from the low values of the Landau pole that we found. Indeed if  one assumes no perturbativity breakdown below the SUSY breaking scale, then
 the Landau pole provides
 an upper limit on $\sqrt{F}$. Interestingly the region that accommodates the Higgs mass, and in particular that  with least tuning
 (i.e., $\kappa > \lambda / \sin 2 \beta$),
 has $\sqrt{F}$ so low that the NLSP should decay within the LHC detectors giving a displaced secondary vertex.

The additional tuning that we discussed above for the region with $\kappa < \lambda / \sin2\beta$ is illustrated by the thinness of the green band in that region compared to the thicker band that we obtain for $\kappa > \lambda/\sin2\beta$. An interesting comment here is that the region with small $\kappa$ is disfavored by the requirement for the realistic vacuum to be a global minimum of the tree-level potential.
Meanwhile for larger $\kappa$ the realistic vacuum tends to be deeper than the other minima.
This matches with our expectation from the analytic study in Section~{\ref{extraminima}} that larger $\lambda$ tends to make  unrealistic minima deeper, while larger $\kappa$ may help the desired minimum to be a global one. As we can see from the pink lines in the figure, in the region preferred by LHC discovery of SM-like Higgs boson, the tuning of the mixing quantified by eq.~(\ref{tuning})  is at the level of 1 part in 5,  which is a mild one.

\bigskip

The left panel in Figure~\ref{pulldownplots} illustrates the situation for {\em sizable} $A$-terms. As in the other figures the green band in the plot is the region where the SM-like Higgs mass is in the range 110~-~125~GeV. The region of the $(\lambda,\, \kappa)$ plane at the left of the purple dashed line is compatible with perturbative unification~\footnote{More precisely the purple dashed line is the boundary of the region that contains couplings that stay perturbative (less than $\sqrt{2\pi}$) up to $\Lambda_{GUT}=1.6\cdot10^{16}$ GeV according to a one-loop RGE with
boundary conditions at 1 TeV and 4 sets of $5+\bar{5}$ of SU(5) entering in the RGE at 5 TeV.}.

Apparently from the figure, there is a tension in accommodating the observed Higgs boson mass in the region compatible with perturbative unification. In fact if one takes literally the boundary for the perturbativity up to the GUT scale there is no point where the tree-level Higgs boson mass is in the green band, that is to say at least 110 GeV. To minimize this tension it is necessary to tune $A_{\lambda}$ in order to saturate the upper bound on the Higgs mass in eq.~(\ref{withA}). This is indicated by the large values of $\Deltamix$ (the pink solid lines in the figure) which shows that the region compatible with perturbative unification has tuning much worse than 1 part in 40. As discussed already for the push-up case one should think the perturbativity boundary as a loose boundary due to the inherent uncertainties in defining a Landau pole. However we do not expect the uncertainties in the definition of the perturbativity boundary to reduce significantly the observed difficulty  to get the Higgs mass with modest loop corrections and perturbative couplings up to the GUT scale.
%

We also remark that the region closer to being compatible with perturbative unification has a very narrow range of  $\kappa$ around 0.09, which suggests that the phenomenology of this region of the NMSSM parameter space will be largely dictated by the requirement to have such a heavy Higgs boson.
Note that in all the regions where the Higgs mass is accommodated, the realistic minimum is a global minimum.

Finally, if we abandon the requirement of perturbative unification, then
we can get the correct Higgs mass with much smaller fine-tuning. For example, for the choice of parameters in the figure, the tuning can be as low as 1 part in 10 when $\lambda \gtrsim 0.7$.

\section{Conclusions}

The LHC has very recently discovered what looks like a SM-like Higgs boson of mass
125 GeV. It is then intriguing to study implications of this observation for extensions of the SM, in particular,
the most popular one of SUSY.
It is well-known that it is difficult to accommodate such a Higgs mass in the minimal version of SUSY (MSSM)
since the tree-level prediction for it is at most $m_Z$ and loop corrections do not suffice to
make up the difference from 125 GeV, unless we sacrifice naturalness of the EW scale.
In this paper, we focused on a specific beyond-the-MSSM scenario, namely, the scale-invariant NMSSM where dimensionful terms in the superpotential can be forbidden by a $Z_3$ symmetry and $\mu$-term for the Higgs doublets can be dynamically generated from the VEV of a singlet
coupled to Higgs doublets.
As a bonus of this singlet-Higgs doublet
coupling, we get an extra quartic coupling for the Higgs doublets, which raises the tree-level SM-like Higgs mass
beyond the $m_Z$ value of the MSSM. Combined with a loop correction from stops of natural light masses
it naively looks that the NMSSM can easily accommodate 125~GeV SM-like Higgs mass.

However, we emphasized that there is a limitation. The above singlet-Higgs doublet coupling along with the VEVs inevitably results in mass terms mixing the singlet and Higgs doublet. Such mixing modifies the mass of SM-like Higgs from the above expectation,
 and the effect does not decouple even if the singlet is heavy. We performed a systematic study of this effect, in particular, we considered both the cases where the SM-like Higgs mass is reduced (``pull-down")
and raised (``push-up")
as a result of the mixing. Also, we carefully distinguished between the cases with negligible vs. non-negligible $A$-terms
and similarly, couplings being perturbative up to GUT scale vs. having Landau poles below. For simplicity and naturalness we chose stop mass below 500~GeV with small stop mixing.

One of our sharp results is that the perturbative case which can preserve the
merit of gauge coupling unification of the MSSM
requires $A_\lambda$ to be tuned with the $\mu$-term in order to
reach 125~GeV for the SM-like Higgs mass.
We quantified this tuning and found that the best attainable tuning is roughly 20\% for push-up and significantly worse tuning for pull-down.
The non-perturbative option allows more room, for example in this case $A$-terms are not necessary. However we showed that there are still non-trivial constraints to be satisfied in order to  achieve the goal of correct SM-like Higgs mass. The most natural region of the non-perturbative parameter space is the one with $\kappa\gtrsim\lambda$ where tuning can be 20\% and the Landau pole scale is typically below $10^{4}$~TeV.
In general, our findings motivate further studies of UV models which lead to desired relations among the parameters of the NMSSM. In particular, in the case of minimal gauge mediated SUSY breaking, $A$-terms are predicted to be small so that the perturbative case is in trouble.

Another highlight of our work is the careful consideration of constraints from global vacuum stability on model parameter space. 
%
%
%
Our analysis shows this constraint can exclude significant portion of parameter space which is allowed otherwise. In particular, this danger of getting deeper unrealistic vacuum can be more relevant in the case of large $\lambda$ associated with low Landau pole.

\bigskip

Finally we comment on some open issues which we did not elaborate in
our current study.
The current Higgs data suggests that the 125~GeV boson is roughly SM-like. We checked that in a sizable region of the parameter space we get a SM-like Higgs boson. In other regions of parameters space compatible with a 125~GeV Higgs it is possible to have modified rates for Higgs decays into SM states or decays into new states as neutralinos and light pseudo-scalars. However we did not investigate these possibilities in details, 
 partly because the data on Higgs decay rates still has  quite large error bars.

There are also neutral scalars and pseudo-scalars
which could potentially be seen in the future at the LHC, although we have estimated
that currently there is no significant constraint here. Another example of an aspect that we did not focus on is the
charged Higgs phenomenology, in particular, it contributes via loop along with top quark to $b \rightarrow s \gamma$. However,
this process also gets a contribution from chargino-stop loop, which may cancel the charged Higgs loop depending on parameters beyond the Higgs sector of the NMSSM, such as Wino and stop masses. Hence, we chose not to pursue this analysis.
Finally, we did not quantify precisely the tuning of the $Z$ mass due to stop contribution to $m_{ H_u }^2$ as it is independent of that required to raise the tree-level mass of SM-like Higgs that we focused on.

\bigskip

The discovery of a SM Higgs-like particle at 125~GeV has strong implications for scale-invariant NMSSM. We have performed a systematic analysis to explore realistic natural regions of the parameter space where such Higgs boson can live.
We find that the inhabitable space shrinks to a few well separated islands. This maps out  well defined directions for future work both for UV model building and phenomenological studies.

\subsubsection*{Note Added}
While this paper was being finalized, we learned that related ideas are being pursued independently by the authors of Ref.~\cite{Gherghetta}.

\section*{Acknowledgments}

We would like to thank  Zacharia Chacko, Ulrich Ellwanger, Tony Gherghetta, Gino Isidori,
Andrey Katz, Riccardo Rattazzi, Daniel Stolarski, Raman Sundrum and Lian-Tao Wang for comments and discussions.
This work is supported in part by NSF Grant No. PHY-0968854. Y.C. and R.F. are supported in part by the Maryland Center for Fundamental Physics and by the NSF Grant No. PHY-0910467. The three of us acknowledge the hospitality of the Aspen Center for Physics, which is supported by the NSF Grant No. PHY-1066293. R.F. thanks CERN TH division for hospitality and support while this work was completed.


\begin{thebibliography}{99}



\bibitem{lhc}
{The CMS Collaboration}, ``{Observation of a new boson at a mass of 125 GeV
  with the CMS experiment at the LHC},''
  \href{http://arxiv.org/abs/1207.7235}{{\ttfamily arXiv:1207.7235 [hep-ex]}}; 
{The ATLAS Collaboration}, ``{Observation of a new particle in the search for
  the Standard Model Higgs boson with the ATLAS detector at the LHC},''
    \href{http://arxiv.org/abs/1207.7214}{{\ttfamily arXiv:1207.7214 [hep-ex]}}.

\bibitem{Draper:2011aa}
P.~Draper, P.~Meade, M.~Reece, and D.~Shih, ``{Implications of a 125 GeV Higgs
  for the MSSM and Low-Scale SUSY Breaking},''
  \href{http://dx.doi.org/10.1103/PhysRevD.85.095007}{{\em Phys.Rev.}
  {\bfseries D85} (2012) 095007},
\href{http://arxiv.org/abs/1112.3068}{{\ttfamily arXiv:1112.3068 [hep-ph]}}.

\bibitem{Hall:2011aa}
L.~J. {Hall}, D.~{Pinner}, and J.~T. {Ruderman}, ``{A Natural SUSY Higgs Near
  125 GeV},''
  \href{http://arxiv.org/abs/1112.2703}{{\ttfamily arXiv:1112.2703 [hep-ph]}}.
  
%
  

\bibitem{Carena:2012zr}
M.~{Carena}, S.~{Gori}, N.~R. {Shah}, and C.~E.~M. {Wagner}, ``{A 125 GeV
  SM-like Higgs in the MSSM and the {$\gamma$}{$\gamma$} rate},''
  \href{http://dx.doi.org/10.1007/JHEP03(2012)014}{{\em Journal of High Energy
  Physics} {\bfseries 3} (Mar., 2012) 14},
  \href{http://arxiv.org/abs/1112.3336}{{\ttfamily arXiv:1112.3336 [hep-ph]}}.

\bibitem{Wymant:2012vl}
C.~{Wymant}, ``{Optimising Stop Naturalness},'' {\em ArXiv e-prints} (Aug.,
  2012) , \href{http://arxiv.org/abs/1208.1737}{{\ttfamily arXiv:1208.1737
  [hep-ph]}}.





\bibitem{Giudice:2006sp}
G.~F. {Giudice} and R.~{Rattazzi}, ``{Living dangerously with low-energy
  supersymmetry},''
  \href{http://dx.doi.org/10.1016/j.nuclphysb.2006.07.031}{{\em Nuclear Physics
  B} {\bfseries 757} (Nov., 2006) 19--46},
  \href{http://arxiv.org/abs/arXiv:hep-ph/0606105}{{\ttfamily
  arXiv:hep-ph/0606105}}.


\bibitem{Dvali:1996ij}
G.~{Dvali}, G.~F. {Giudice}, and A.~{Pomarol}, ``{The {$\mu$}-problem in
  theories with gauge-mediated supersymmetry breaking},''
  \href{http://dx.doi.org/10.1016/0550-3213(96)00404-X}{{\em Nuclear Physics B}
  {\bfseries 478} (Feb., 1996) 31--45},
  \href{http://arxiv.org/abs/arXiv:hep-ph/9603238}{{\ttfamily
  arXiv:hep-ph/9603238}}.

\bibitem{Csaki:2008dp}
  C.~Csaki, A.~Falkowski, Y.~Nomura and T.~Volansky,
  ``New Approach to the mu-Bmu Problem of Gauge-Mediated Supersymmetry Breaking,''
  Phys.\ Rev.\ Lett.\  {\bf 102}, 111801 (2009)
   \href{http://arxiv.org/abs/0809.4492}{{\ttfamily arXiv:0809.4492 [hep-ph]}}.

\bibitem{de-Simone:2011uq}
A.~{de Simone}, R.~{Franceschini}, G.~F. {Giudice}, D.~{Pappadopulo}, and
  R.~{Rattazzi}, ``{Lopsided gauge mediation},''
  \href{http://dx.doi.org/10.1007/JHEP05(2011)112}{{\em Journal of High Energy
  Physics} {\bfseries 5} (May, 2011) 112},
  \href{http://arxiv.org/abs/1103.6033}{{\ttfamily arXiv:1103.6033 [hep-ph]}}.

\bibitem{Barbieri:1987fn}
R.~Barbieri and G.~Giudice, ``{Upper Bounds on Supersymmetric Particle
  Masses},''
\href{http://dx.doi.org/10.1016/0550-3213(88)90171-X}{{\em Nucl.Phys.}
  {\bfseries B306} (1988) 63}.

\bibitem{The-DELPHI-Collaboration:-J.Abdallah:2003rr}
{The DELPHI Collaboration: J.~Abdallah}, ``{Searches for supersymmetric
  particles in e+e- collisions up to 208 GeV and interpretation of the results
  within the MSSM},'' {\em ArXiv High Energy Physics - Experiment e-prints}
  (Nov., 2003) , \href{http://arxiv.org/abs/arXiv:hep-ex/0311019}{{\ttfamily
  arXiv:hep-ex/0311019}}.



\bibitem{Ellwanger:2009dp}
U.~{Ellwanger}, C.~{Hugonie}, and A.~M. {Teixeira}, ``{The Next-to-Minimal
  Supersymmetric Standard Model},''
  \href{http://dx.doi.org/10.1016/j.physrep.2010.07.001}{{\em Phys.Rept.}
  {\bfseries 496} (Nov., 2010) 1--77},
  \href{http://arxiv.org/abs/0910.1785}{{\ttfamily arXiv:0910.1785 [hep-ph]}}.

  \bibitem{Abel:1995wk}
S.~A. {Abel}, S.~{Sarkar}, and P.~L. {White}, ``{On the cosmological domain
  wall problem for the minimally extended supersymmetric standard model},''
  \href{http://dx.doi.org/10.1016/0550-3213(95)00483-9}{{\em Nuclear Physics B}
  {\bfseries 454} (Feb., 1995) 663--681},
  \href{http://arxiv.org/abs/arXiv:hep-ph/9506359}{{\ttfamily
  arXiv:hep-ph/9506359}}.








\bibitem{Abel:1996fk}
S.~A. {Abel}, ``{Destabilising divergences in the NMSSM},''
  \href{http://dx.doi.org/10.1016/S0550-3213(96)00470-1}{{\em Nuclear Physics
  B} {\bfseries 480} (Feb., 1996) 55--72},
  \href{http://arxiv.org/abs/arXiv:hep-ph/9609323}{{\ttfamily
  arXiv:hep-ph/9609323}}.




\bibitem{Panagiotakopoulos:1998yw}
C.~{Panagiotakopoulos} and K.~{Tamvakis}, ``{Stabilized NMSSM without domain
  walls},'' \href{http://dx.doi.org/10.1016/S0370-2693(98)01493-2}{{\em Physics
  Letters B} {\bfseries 446} (Jan., 1999) 224--227},
  \href{http://arxiv.org/abs/arXiv:hep-ph/9809475}{{\ttfamily
  arXiv:hep-ph/9809475}}.




\bibitem{Ellwanger:2007kx}
U.~{Ellwanger} and C.~{Hugonie}, ``{The Upper Bound on the Lightest Higgs Mass
  in the NMSSM Revisited},''
  \href{http://dx.doi.org/10.1142/S0217732307023870}{{\em Modern Physics
  Letters A} {\bfseries 22} (2007) 1581--1590},
  \href{http://arxiv.org/abs/arXiv:hep-ph/0612133}{{\ttfamily
  arXiv:hep-ph/0612133}}.

\bibitem{Barbieri:2006bg}
R.~{Barbieri}, L.~J. {Hall}, Y.~{Nomura}, and V.~S. {Rychkov}, ``{Supersymmetry
  without a light Higgs boson},''
  \href{http://dx.doi.org/10.1103/PhysRevD.75.035007}{{\em Phys. Rev. D}
  {\bfseries 75} no.~3, (Feb., 2007) 035007},
  \href{http://arxiv.org/abs/arXiv:hep-ph/0607332}{{\ttfamily
  arXiv:hep-ph/0607332}}.

\bibitem{Barbieri:2008yq}
R.~{Barbieri}, D.~{Pappadopulo}, V.~S. {Rychkov}, L.~J. {Hall}, and A.~Y.
  {Papaioannou}, ``{An alternative NMSSM phenomenology with manifest
  perturbative unification},''
  \href{http://dx.doi.org/10.1088/1126-6708/2008/03/005}{{\em Journal of High
  Energy Physics} {\bfseries 3} (Mar., 2008) 5},
  \href{http://arxiv.org/abs/0712.2903}{{\ttfamily arXiv:0712.2903 [hep-ph]}}.




\bibitem{Masip:1998zr}
M.~{Masip}, R.~{Mu{\~n}oz-Tapia}, and A.~{Pomarol}, ``{Limits on the mass of
  the lightest Higgs boson in supersymmetric models},''
  \href{http://dx.doi.org/10.1103/PhysRevD.57.R5340}{{\em \prd} {\bfseries 57}
  (May, 1998) 5340},
  \href{http://arxiv.org/abs/arXiv:hep-ph/9801437}{{\ttfamily
  arXiv:hep-ph/9801437}}.

\bibitem{Schael:2006cr}
S.~{Schael},   et~al., ``{Search for neutral MSSM Higgs bosons at LEP},''
  \href{http://dx.doi.org/10.1140/epjc/s2006-02569-7}{{\em European Physical
  Journal C} {\bfseries 47} (Sept., 2006) 547--587},
  \href{http://arxiv.org/abs/arXiv:hep-ex/0602042}{{\ttfamily
  arXiv:hep-ex/0602042}}.

\bibitem{Harnik:2004fj}
R.~{Harnik}, G.~D. {Kribs}, D.~T. {Larson}, and H.~{Murayama}, ``{Minimal
  supersymmetric fat Higgs model},''
  \href{http://dx.doi.org/10.1103/PhysRevD.70.015002}{{\em \prd} {\bfseries 70}
  no.~1, (July, 2004) 015002},
  \href{http://arxiv.org/abs/arXiv:hep-ph/0311349}{{\ttfamily
  arXiv:hep-ph/0311349}}.

 \bibitem{Birkedal:2004uq}
A.~Birkedal, Z.~Chacko, and Y.~Nomura, ``{Relaxing the upper bound on the mass
  of the lightest supersymmetric Higgs boson},''
  \href{http://dx.doi.org/10.1103/PhysRevD.71.015006}{{\em Phys.Rev.}
  {\bfseries D71} (2005) 015006},
\href{http://arxiv.org/abs/hep-ph/0408329}{{\ttfamily arXiv:hep-ph/0408329
  [hep-ph]}}.



\bibitem{Hardy:2012fk}
E.~{Hardy}, J.~{March-Russell}, and J.~{Unwin}, ``{Precision Unification in
  $\lambda$SUSY with a 125 GeV Higgs},'' {\em ArXiv e-prints} (July,
  2012) , \href{http://arxiv.org/abs/1207.1435}{{\ttfamily arXiv:1207.1435
  [hep-ph]}}.
  
  
\bibitem{Ross:2012nr} 
  G.~G.~Ross, K.~Schmidt-Hoberg and F.~Staub,
  ``The Generalised NMSSM at One Loop: Fine Tuning and Phenomenology,''
  JHEP {\bf 1208}, 074 (2012)
  [arXiv:1205.1509 [hep-ph]].






\bibitem{Kang:2012qf}
Z.~{Kang}, J.~{Li}, and T.~{Li}, ``{On Naturalness of the (N)MSSM},''
  \href{http://arxiv.org/abs/1201.5305}{{\ttfamily arXiv:1201.5305 [hep-ph]}}.


\bibitem{Arvanitaki:2011ck}
A.~{Arvanitaki} and G.~{Villadoro}, ``{A non Standard Model Higgs at the LHC as
  a sign of naturalness},''
  \href{http://dx.doi.org/10.1007/JHEP02(2012)144}{{\em Journal of High Energy
  Physics} {\bfseries 2} (Feb., 2012) 144},
  \href{http://arxiv.org/abs/1112.4835}{{\ttfamily arXiv:1112.4835 [hep-ph]}}.

\bibitem{Ellwanger:2011aa}
U.~{Ellwanger}, ``{A Higgs boson near 125 GeV with enhanced di-photon signal in
  the NMSSM},'' \href{http://dx.doi.org/10.1007/JHEP03(2012)044}{{\em Journal
  of High Energy Physics} {\bfseries 3} (Mar., 2012) 44},
  \href{http://arxiv.org/abs/1112.3548}{{\ttfamily arXiv:1112.3548 [hep-ph]}}.

\bibitem{Ellwanger:2012ad}
U.~{Ellwanger} and C.~{Hugonie}, ``{Higgs bosons near 125 GeV in the NMSSM with
  constraints at the GUT scale},'' {\em ArXiv e-prints} (Mar., 2012) ,
  \href{http://arxiv.org/abs/1203.5048}{{\ttfamily arXiv:1203.5048 [hep-ph]}}.

\bibitem{Cao:2012hb}
J.~{Cao}, Z.~{Heng}, J.~M. {Yang}, Y.~{Zhang}, and J.~{Zhu}, ``{A SM-like Higgs
  near 125 GeV in low energy SUSY: a comparative study for MSSM and NMSSM},''
  \href{http://dx.doi.org/10.1007/JHEP03(2012)086}{{\em Journal of High Energy
  Physics} {\bfseries 3} (Mar., 2012) 86},
  \href{http://arxiv.org/abs/1202.5821}{{\ttfamily arXiv:1202.5821 [hep-ph]}}.

\bibitem{Cao:2012sf}
J.~{Cao}, Z.~{Heng}, J.~M. {Yang}, and J.~{Zhu}, ``{Status of low energy SUSY
  models confronted with the LHC 125 GeV Higgs data},'' {\em ArXiv e-prints}
  (July, 2012) , \href{http://arxiv.org/abs/1207.3698}{{\ttfamily
  arXiv:1207.3698 [hep-ph]}}.

\bibitem{Albornoz-Vasquez:2012yf}
D.~{Albornoz Vasquez}, G.~{Belanger}, C.~{Boehm}, J.~{Da Silva},
  P.~{Richardson}, and C.~{Wymant}, ``{The 125 GeV Higgs in the NMSSM in light
  of LHC results and astrophysics constraints},'' {\em ArXiv e-prints} (Mar.,
  2012) , \href{http://arxiv.org/abs/1203.3446}{{\ttfamily arXiv:1203.3446
  [hep-ph]}}.


\bibitem{Romao:1986jy}
J.~Romao, ``{Spontaneous CP violation in SUSY models: a no-go theorem},''
\href{http://dx.doi.org/10.1016/0370-2693(86)90522-8}{{\em Phys.Lett.}
  {\bfseries B173} (1986) 309}.












\bibitem{Papucci:2011wy} 
  M.~Papucci, J.~T.~Ruderman and A.~Weiler,
  JHEP {\bf 1209}, 035 (2012)
  [arXiv:1110.6926 [hep-ph]].


\bibitem{Degrassi:2010lr}
G.~{Degrassi} and P.~{Slavich}, ``{On the radiative corrections to the neutral
  Higgs boson masses in the NMSSM},''
  \href{http://dx.doi.org/10.1016/j.nuclphysb.2009.09.018}{{\em Nuclear Physics
  B} {\bfseries 825} (Jan., 2010) 119--150},
  \href{http://arxiv.org/abs/0907.4682}{{\ttfamily arXiv:0907.4682 [hep-ph]}}.













\bibitem{The-BABAR-Collaboration:2012qf}
{The BABAR Collaboration}, ``{Measurement of $B(B \to X_s \gamma)$, the $B\to
  X_s \gamma$ photon energy spectrum, and the direct CP asymmetry in $B \to
  X_{s+d} \, \gamma$ decays},''
  \href{http://arxiv.org/abs/1207.5772}{{\ttfamily arXiv:1207.5772 [hep-ex]}}.


  \bibitem{Kanehata:2011gf}
Y.~{Kanehata}, T.~{Kobayashi}, Y.~{Konishi}, O.~{Seto}, and T.~{Shimomura},
  ``{Constraints from Unrealistic Vacua in the Next-to-Minimal Supersymmetric
  Standard Model},'' {\em Progress of Theoretical Physics} {\bfseries 126}
  (Dec., 2011) 1051--1076, \href{http://arxiv.org/abs/1103.5109}{{\ttfamily
  arXiv:1103.5109 [hep-ph]}}.

\bibitem{Kobayashi:2012fj}
T.~{Kobayashi}, T.~{Shimomura}, and T.~{Takahashi}, ``{Constraining the Higgs
  sector from False Vacua in the Next-to-Minimal Supersymmetric Standard
  Model},''
  \href{http://arxiv.org/abs/1203.4328}{{\ttfamily arXiv:1203.4328 [hep-ph]}}.




\bibitem{nmssmtools}
U.~Ellwanger, J.~F.~Gunion, C.~Hugonie \url{http://www.th.u-psud.fr/NMHDECAY/nmssmtools.html}\,.
%

 \bibitem{Jeong:2012fk}
K.~S. {Jeong}, Y.~{Shoji}, and M.~{Yamaguchi}, ``{Singlet-doublet Higgs mixing
  and its implications on the Higgs mass in the PQ-NMSSM},''   \href{http://arxiv.org/abs/1205.2486}{{\ttfamily
  arXiv:1205.2486 [hep-ph]}}.





\bibitem{Larsen:2012ys}
G.~{Larsen}, Y.~{Nomura}, and H.~L.~L. {Roberts}, ``{Supersymmetry with Light
  Stops},''
  \href{http://arxiv.org/abs/1202.6339}{{\ttfamily arXiv:1202.6339 [hep-ph]}}.



\bibitem{Chang:2004db}
S.~{Chang}, C.~{Kilic}, and R.~{Mahbubani}, ``{New fat Higgs: Increasing the
  MSSM Higgs mass with natural gauge unification},''
  \href{http://dx.doi.org/10.1103/PhysRevD.71.015003}{{\em Phys. Rev. D}
  {\bfseries 71} no.~1, (Jan., 2005) 015003},
  \href{http://arxiv.org/abs/arXiv:hep-ph/0405267}{{\ttfamily
  arXiv:hep-ph/0405267}}.

%


\bibitem{Gherghetta}
T.~{Gherghetta}, B.~{von Harling}, A.~D. {Medina}, and M.~A. {Schmidt}, ``{The
  Scale-Invariant NMSSM and the 126 GeV Higgs Boson},''
  (Dec., 2012) , \href{http://arxiv.org/abs/1212.5243}{{\ttfamily
  arXiv:1212.5243 [hep-ph]}}.

\end{thebibliography}
\end{document}